\documentclass[pra,twocolumn,superscriptaddress,showpacs,nofootinbibfloatfix,amsmath,amsfonts,amssymb]{revtex4-1}%
\usepackage{amsmath,amsfonts,amssymb,color}
\usepackage{amsthm}
\usepackage{leftidx}
\usepackage{graphicx}
\usepackage{xcolor}
\usepackage{dcolumn}
\usepackage{bm}
\usepackage{epstopdf}
\usepackage{epsfig}
\usepackage{environ}

\usepackage{float}
\usepackage[T1]{fontenc}
\usepackage[latin9]{inputenc}
\usepackage{setspace}
\usepackage{esint}

\newcommand{\n}{\noindent}

\begin{document}
	\title{{Coupled-wire construction of static and Floquet second-order topological insulators}}
%\title{Constructing static and Floquet second-order topological insulators by stacking one-dimensional topological insulators}
	\author{Raditya Weda Bomantara}
	\email{phyrwb@nus.edu.sg}
	\affiliation{%
		Department of Physics, National University of Singapore, Singapore 117543
	}
	\author{Longwen Zhou}
	\email{zhoulw13@u.nus.edu}
	\affiliation{%
		Department of Physics, College of Information Science and Engineering, Ocean University of China, Qingdao, China 266100
	}
	%\affiliation{%
	%Department of Physics, National University of Singapore, Singapore 117543
	%}
	\author{{Jiaxin Pan}}
	\affiliation{%
	Department of Physics, College of Information Science and Engineering, Ocean University of China, Qingdao, China 266100
	}
	\author{Jiangbin Gong}%
	\email{phygj@nus.edu.sg}
	\affiliation{%
		Department of Physics, National University of Singapore, Singapore 117543
	}
	\date{\today}
	
	%%%%%%%%%%%%%%%%%%%% ABSTRACT %%%%%%%%%%%%%%%%%%%%%%%%
	%\begin{linenumbers}
	
	\vspace{2cm}
	
	\begin{abstract}
	%In recent years, second-order topological insulators (SOTI) which exhibit gapless boundary states at their hinges or corners have been extensively studied. 
	{Second-order topological insulators (SOTI) exhibit protected gapless boundary states at their hinges or corners.}
	%In this paper, we report a means to construct such SOTIs in static and Floquet systems by stacking one-dimensional topological insulators and coupling them with dimerized hopping amplitude in the spirit of the Su-Schrieffer-Heeger model, which is protected solely by chiral symmetry and does not require the presence of any spatial symmetry. 
	{In this paper, we propose a generic means to construct SOTIs in static and Floquet systems by coupling one-dimensional topological insulator wires along a second dimension through dimerized hopping amplitudes.}
	{The Hamiltonian of such SOTIs admits a Kronecker sum structure, making it possible for obtaining its features by analyzing two constituent one-dimensional lattice Hamiltonians defined separately in two orthogonal dimensions.}
	{The resulting topological corner states do not rely on any delicate spatial symmetries, but are solely protected by the chiral symmetry of the system. We further utilize our idea to construct Floquet SOTIs, whose number of topological corner states is arbitrarily tunable via changing the hopping amplitudes of the system. Finally, we propose to detect the topological invariants of static and Floquet SOTIs constructed with our approach in experiments by measuring the mean chiral displacements of wavepackets.}
	%We find that the Hamiltonian describing such SOTI admits a Kronecker sum structure between two one-dimensional (1D) Hamiltonian describing the coupling in the $x$- and $y$-directions, so that most of their features can be obtained by analyzing each of these 1D Hamiltonian separately. Finally, we utilize this idea to construct a Floquet SOTI which is capable of hosting arbitrarily many topological corner modes, tunable via some of the system parameters.

	\end{abstract}
	%\pacs{03.65.Vf, 05.60.Gg, 05.30.Rt, 73.20.At}
	
	\maketitle
	
	\section{Introduction}
	
	Topological phases of matter have emerged as an active research topic studied by both theorists and experimentalists since the last decade. As the name suggests, such phases of matter are characterized by the topology of their bulk states, the latter of which manifests itself as physical observables at their boundaries. For example,  {quantum spin Hall insulators} can be distinguished from normal insulators by the value of the $\mathbb{Z}_2$ topological invariant that their bulk states possess, which determines the presence or absence of the topologically protected helical edge states at the boundaries of the systems~\cite{TI,TI2,TI3}. The edge properties of topological phases are thus robust to local perturbations that preserve their topology as well as the symmetries protecting them. Consequently, topological phases are considered as a promising platform for designing robust electronic/spintronic devices, offering (almost) dissipationless and faster charge transfers \cite{TIA}, as well as providing protections at the hardware level in the realization of fault-tolerant quantum computations~\cite{TIA2}.
	
	In recent years, a new type of topological phases whose topology manifests itself at the boundaries of their boundaries has been discovered and termed higher-order topological phases \cite{HTI0,HTI1,HTI2,HTI3,HTI4,HTI5,HTI6,HTI7,HTI8,HTI9,HTI10,HTI11,HTI12,HTI13,HTI14,HTI15,HTI16,HTI17,HTI18,HTI19,HTI20,LU,HTI21,HTI22,HTI23}. In particular, a $d$-dimensional $n$th-order topological insulator (where $d\geq n$) is characterized by the existence of topologically protected ($d-n$)-dimensional boundary states and gapped higher-dimensional boundary and bulk states. 
	
	%Unlike first-order topological phases, which can exist even in the absence of any symmetries, most of the existing proposals on the construction of higher-order topological phases relies on the presence of additional spatial (reflection, inversion, or rotational) symmetries. 
	
	{  Most of the existing proposals of higher-order topological phases relies on the presence of spatial (reflection, inversion, and/or rotational) symmetries. By contrast, it is well known that first-order topological phases may exist even in the absence of these spatial symmetries, which can be further characterized solely by the presence of time-reversal, particle-hole, and chiral symmetries through the Altland-Zirnbauer (AZ) classification scheme \cite{AZ}.} It thus raises a fundamental question regarding the existence of higher-order topological phases in the absence of any spatial symmetries, which may provide further insight into the similarity between first- and higher-order topological phases. This question has been explored recently in Ref.~\cite{LU}, which proposes the construction of higher-order topological insulators in square and cubic lattices by coupling together four and eight Su-Schrieffer-Heeger (SSH) systems \cite{SSH}, respectively. Each of them describes a one-dimensional (1D) topological insulating model characterized by a topological winding number that determines the presence or absence of zero energy states at each end of the system. By construction, such models are protected solely by chiral symmetry without the need for additional spatial symmetries. However, it remained an open question if 
	%four SSH systems are really necessary to construct higher-order topological insulators without any spatial symmetries, and if 
	a more general construction based on 1D topological insulating models other than the SSH model is possible. %{Furthermore, to the best of our knowledge, no extensions of SOTIs to Floquet systems (i.e., systems that are driven periodically in time) have ever been made.}
	
	In this paper, we propose a general framework for constructing second-order topological insulators (SOTI) in a square lattice by means of coupling an array of 1D topological insulators with dimerized inter-array hopping amplitudes, as illustrated in Fig.~\ref{stack}. As will be shown below, the total Hamiltonian of such a system can be written as a Kronecker sum of two 1D topological insulating Hamiltonians, enabling one to characterize the topology of the full system from that of its 1D Hamiltonian constituents separately. In particular, we show that topological corner modes exist only if both 1D Hamiltonian constituents are topologically nontrivial, which persist even in the presence of perturbations breaking all but the chiral symmetry, as well as small perturbations breaking the Kronecker sum structure of the system. 
	
	\begin{figure}
		\begin{center}
			\includegraphics[scale=0.45]{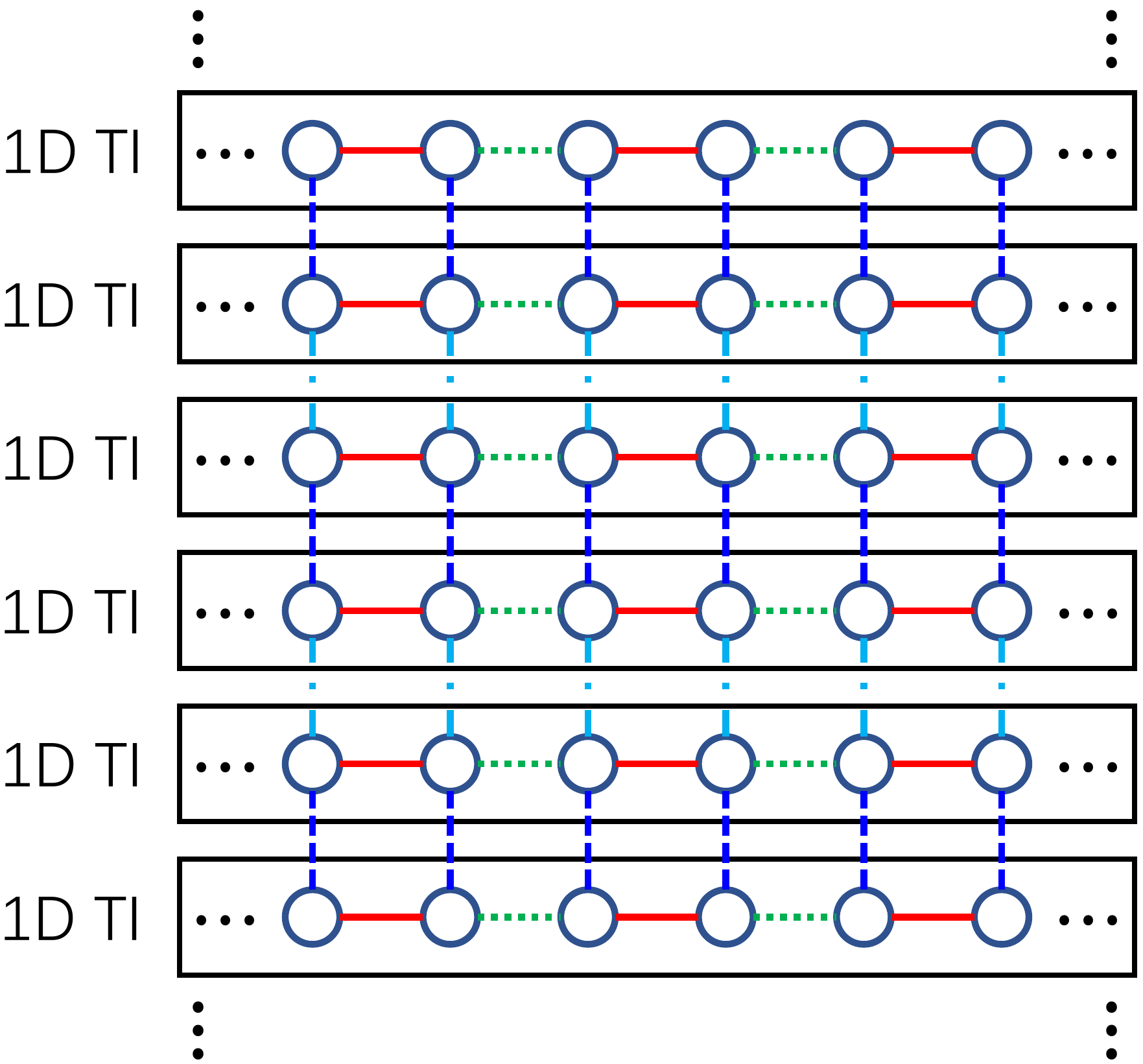}
		\end{center}
		\caption{
			%Constructing SOTI by stacking an array of 1D topological insulators (such as SSH model as shown in the inset). Red and blue lines denote different inter-array coupling strength, whereas purple and orange lines depict the different coupling strength between pairs of lattice sites within each array.
		{Constructing SOTI by stacking an array of 1D topological insulators (such as SSH model as shown in each longitudinal box). Each circle represents a lattice site. Red solid and green dotted lines denote different coupling strength between pairs of lattice sites within each array, whereas blue dashed and cyan dash-dotted lines depict different inter-array coupling strength.}}
		\label{stack} 
	\end{figure}
	
	By the same mechanism outlined above, Floquet (periodically driven) SOTIs can be obtained by coupling an array of 1D Floquet topological insulators with the same (static) dimerized inter-array hopping amplitude. It is noted that the studies of Floquet topological phases have attracted tremendous interest in recent years due to their capability to exhibit properties with no static analogue, such as the existence of {  edge states pinned at quasienergy (the analogue of energy in Floquet systems) $\frac{\pi}{T}$ \cite{cref,DG,RG,RG2,RG3,LW}} and anomalous edge states which do not satisfy the usual bulk-edge correspondence \cite{aes}. While Floquet first-order topological phases have been extensively studied \cite{cref,DG,RG,RG2,RG3,LW,aes,FTP-1,FTP0,FTP1,FTP2,FTP3,FTP4,FTP5,FTP6,FTP7,FTP8,FTP9,FTP10,FTP11,FTP12,FTP13,FTP14,FTP15,FTP16,MCD1,MCD2}, their extension to higher-order topological phases has never been explored so far to our knowledge. 
	
	Through a relatively simple proposal for constructing Floquet SOTI, this paper is thus hoped to motivate future studies to explore some unique opportunities offered by Floquet higher-order topological phases. To that end, we show in this paper how such a Floquet SOTI can accommodate arbitrarily many topological corner modes at both quasienergies zero and $\frac{\pi}{T}$, a feature which cannot be found in any static SOTI. In particular, the coexistence of topological corner modes at quasienergy zero and $\frac{\pi}{T}$ by itself already represents an unforeseen scenario which can be utilized for topological quantum computation \cite{RG7}. The existence of many topological corner modes is also expected to be useful for quantum memory applications at the very least.       
	
	This paper is organized as follows. We introduce our proposal in Sec.~\ref{SSOTI} by starting with an explicit model describing an array of SSH model coupled together with another SSH-like coupling in the $y$-direction and present the analytical expression of the corner modes. In Sec.~\ref{SSOTI2}, we show that the full Hamiltonian of the system can be written as a Kronecker sum of two 1D SSH Hamiltonian. As a result, the latter symmetry and topological properties can be obtained from those of its 1D Hamiltonian constituents separately. In Sec.~\ref{SSOTI3}, we discuss the difference between our proposal and that of Ref.~\cite{LU}, 
	%{ present a possible detection of the winding numbers by means of mean chiral displacement,} 
	and the robustness of our proposal in the presence of small perturbations which destroy the Kronecker sum structure of the full Hamiltonian. In Sec.~\ref{general}, we extend our proposal to construct Floquet SOTI which may host topological corner modes at quasienergy zero and $\frac{\pi}{T}$ (Floquet zero and $\pi$ corner modes). We present an explicit model of such Floquet SOTI in Sec.~\ref{FSOTI} and show how arbitrarily many zero and $\pi$ Floquet corner modes can be systematically obtained by tuning some system parameters. {  In Sec.~\ref{detect}, we propose to detect the bulk topological invariants of static and Floquet SOTIs by measuring the mean chiral displacement of a wavepacket.}
	We summarize our results and discuss some future directions in Sec.~\ref{conc}.
	
	%\section{Constructing static SOTI by stacking 1D topological phases}
	\section{Coupled-wire construction of static SOTI}
	\label{SSOTI}
	
	{In this section, we introduce our scheme of constructing static SOTIs via coupling topological insulator wires, and present explicit model calculations to demonstrate our findings}. 
	
	We start by considering a prototypical tight-binding Hamiltonian $\mathcal{H}$, which describes particles hopping on a 2D lattice:
	\begin{eqnarray}
	\mathcal{H}&=& \sum_{i=1}^{N_x} \sum_{j=1}^{N_y} \left\lbrace \left[J_y + (-1)^j \delta J_y\right] |i,j+1\rangle \langle i,j | \right. \nonumber \\
	&& + \left.\left[J_x + (-1)^i \delta J_x\right] |i+1,j\rangle \langle i,j | +\rm{h.c.}\right\rbrace \;. \label{SSH}
	\end{eqnarray}
	
	\n Here $J_{x(y)}\pm \delta J_{x(y)}$ denote dimerized hopping amplitudes in the $x$-($y$-)direction, $|i,j\rangle$ denotes the basis state at lattice site $(x,y)=(i,j)$, $N_x$ and $N_y$ are the number of lattice sites in $x$- and $y$-directions, respectively. Without loss of generality, we will take $J_x$ and $J_y$ to be nonnegative throughout this paper. Eq.~(\ref{SSH}) can thus be understood as an array of SSH chains along the $x$-direction, coupled with each other by another SSH-type dimerized hopping along the $y$-direction.
	 {Such a model Hamiltonian may be realized experimentally in silicon photonic setups~\cite{SSHSSHExp}.}
	{  Early on, a model similar to that presented above has also been studied in Ref.~\cite{HTI1,HTI2}, and was shown to exhibit corner modes but has a vanishing bulk quadrupole invariant. In the following, we argue that such a model actually qualifies as another type of SOTI, characterized by the robustness of its corner modes, the existence of edge and bulk band gaps, and a different type of bulk topological invariant.}
	
	{  To understand how the above model may host corner modes, we may start by noticing that} if the hopping amplitude and dimerization parameter along $y$-direction satisfy $J_y=\delta J_y=0$, the system described by Hamiltonian $\mathcal{H}$ reduces to $N_y$ identical copies of 1D SSH chain. Each of them can be in either a topologically trivial~($\delta J_x <0$) or a nontrivial~($\delta J_x >0$) phase. In the topologically nontrivial regime, a pair of degenerate zero-energy edge states (also called zero modes) appears at the two ends of each chain, resulting in totally $2N_y$ such degenerate edge states in the whole system. When $J_y,\delta J_y\neq 0$, all these zero modes will in general be coupled together, lifting their degeneracy. However, if $J_y=\delta J_y$, the two pairs of zero modes appearing at the ends of the first ($j=1$) and last ($j=N_y$) arrays will be decoupled, and therefore remaining degenerate. In this case, four zero modes emerge as corner states in the whole system.
	
	{Away from the fully dimerized limit $J_y=\delta J_y$ along $y$-direction, it can be analytically shown {  (as detailed in Appendix~\ref{app})} that there are four corner modes in the system if $\delta J_y>0$ (in addition to $\delta J_x>0$ as required for each SSH chain to host zero edge modes), which are given by}
	
	\begin{equation}
	|0_{(X,Y)}\rangle = \sum_{i=1}^{N_x/2} \sum_{j=1}^{N_y/2} (-1)^{i+j} \left(\frac{\mathcal{J}_y'}{\mathcal{J}_y}\right)^{j-1} \left(\frac{\mathcal{J}_x'}{\mathcal{J}_x}\right)^{i-1} |X_i,Y_j\rangle \;, \label{szero} 
	\end{equation} 
	
	\n where $\mathcal{J}_{x(y)}'=J_{x(y)}-\delta J_{x(y)}$, $\mathcal{J}_{x(y)}=J_{x(y)}+\delta J_{x(y)}$, $X=1,N_x+2$, $Y=1,N_y+2$, $X_i=|X-2i|$, and $Y_j=|Y-2j|$. {  For a finite lattice, applying $\mathcal{H}$ to $|0_{(X,Y)}\rangle$ directly results in terms proportional to $\left(\mathcal{J}_{x}'/\mathcal{J}_{x}\right)^{N_x/2+1}$ and/or $\left(\mathcal{J}_{y}'/\mathcal{J}_{y}\right)^{N_y/2+1}$, which become smaller as $N_x$ and $N_y$ are increased, so that $|0_{(X,Y)}\rangle$ can be regarded as approximate zero energy solutions to $\mathcal{H}$.}
	{In the following subsection, we discuss the symmetry protecting these corner modes and introduce a topological invariant to characterize them.}
	\vspace{0.5cm}
	
	\subsection{Symmetry analysis and topological invariant}
	\label{SSOTI2}
	
	Under periodic boundary conditions (PBC), the Hamiltonian $\mathcal{H}$ in Eq.~(\ref{SSH}) can be rewritten in momentum space as
	\begin{eqnarray}
	\mathcal{H} &=& \sum_{k_x,k_y} |k_x,k_y\rangle h(k_x,k_y)\langle k_x,k_y| \nonumber \\
	&=& \sum_{k_x,k_y} |k_x,k_y\rangle h_{x,k}\oplus h_{y,k}\langle k_x,k_y| \;, \nonumber \\
	h_{S,k} &=& h_{a,S} \sigma^{(S)}_x +h_{b,S} \sigma^{(S)}_y \;, \label{mstat} 
	\end{eqnarray}
	
	\n where $S=x,y$, $h_{a,S}=\left[ J_S-\delta J_S + \left(J_S+\delta J_S\right) \cos k_S \right]$, $h_{b,S}=\left(J_S + \delta J_S \right) \sin k_S $, $k_S$ and $\sigma^{(S)}$'s are respectively quasimomenta and Pauli matrices acting in the sublattice/pseudospin subspace in the $S$-direction. It is noted that each $h_{S,k}$ is simply the momentum space Hamiltonian describing an SSH model, which possesses inversion, time-reversal, particle-hole, and chiral symmetries respectively dictated by the operators $\mathcal{I}_S=\sigma^{(S)}_x$, $\mathcal{T}_S=\mathcal{K}$, $\mathcal{P}_S=\sigma^{(S)}_z \mathcal{K}$, and $\Gamma_S = \sigma^{(S)}_z$, where $\mathcal{K}$ is the complex conjugation operator. It thus belongs to class BDI in the AZ classification scheme \cite{AZ}, and is characterized by a winding number topological invariant.
	
	Due to the Kronecker sum structure of Eq.~(\ref{mstat}), the full Hamiltonian $\mathcal{H}$ also possesses inversion, time-reversal, particle-hole, and chiral symmetries described by the operators $\mathcal{I}=\sigma^{(x)}_x\sigma^{(y)}_x$, $\mathcal{T}=\mathcal{K}$, $\mathcal{P}=\sigma^{(x)}_z \sigma^{(y)}_z \mathcal{K}$, and $\Gamma = \sigma^{(x)}_z \sigma^{(y)}_z$, {  which satisfy $\mathcal{I}h(k_x,k_y)\mathcal{I}^{-1}=h(-k_x,-k_y)$, $\mathcal{T}h(k_x,k_y)\mathcal{T}^{-1}=h(-k_x,-k_y)$, $\mathcal{P}h(k_x,k_y)\mathcal{P}^{-1}=-h(-k_x,-k_y)$, and $\Gamma h(k_x,k_y)\Gamma=-h(k_x,k_y)$. In addition, the separate inversion symmetries of $h_{x,k}$ and $h_{y,k}$ can also be regarded as two commuting reflection symmetries of the full Hamiltonian $\mathcal{H}$, i.e., $M_x=\mathcal{I}_x$ and $M_y=\mathcal{I}_y$, such that $M_x h(k_x,k_y) M_x^{-1}=h(-k_x,k_y)$ and $M_x h(k_x,k_y) M_x^{-1}=h(k_x,-k_y)$. However, as will be elucidated later on, only the chiral symmetry plays a role in protecting the topological corner modes in our system.
	
	As another consequence of the Kronecker sum structure of $\cal{H}$, the winding number of $h_{x,k}$ ($h_{y,k}$) still dictates the existence of edge states of $\mathcal{H}$ under open boundary conditions (OBC) at the edges of the lattice in the $x$($y$)-direction,} but they are no longer pinned at zero energy since such edge states can be expressed as the tensor product between the edge states of $h_{x,k}$ ($h_{y,k}$) and the bulk states of $h_{y,k}$ ($h_{x,k}$), which therefore have nonzero energies. However, if the winding number of $h_{x,k}$ and $h_{y,k}$ are both nonzero, zero energy eigenstates of $\mathcal{H}$ under OBC can be constructed as a tensor product between the edge states of $h_{x,k}$ and $h_{y,k}$, both having zero energies. By construction, such states are localized at both edges of the lattice and are thus corner modes. Therefore, the existence of corner modes of $\mathcal{H}$ or any Hamiltonian with similar Kronecker sum structures is determined by the product of the topological invariant (e.g. winding number) of each Kronecker sum component.
	
	The winding number associated with the Hamiltonian in the form of Eq.~(\ref{mstat}) is defined as
	\begin{equation}
	\nu_S = \frac{1}{2\pi\mathrm{i}} \int dk_x H_{S,k}^{-1} \frac{d}{d k_x} H_{S,k} \;, \label{wn}   
	\end{equation}
	
\begin{figure*}
	 	\begin{center}
	 		\includegraphics[scale=0.45]{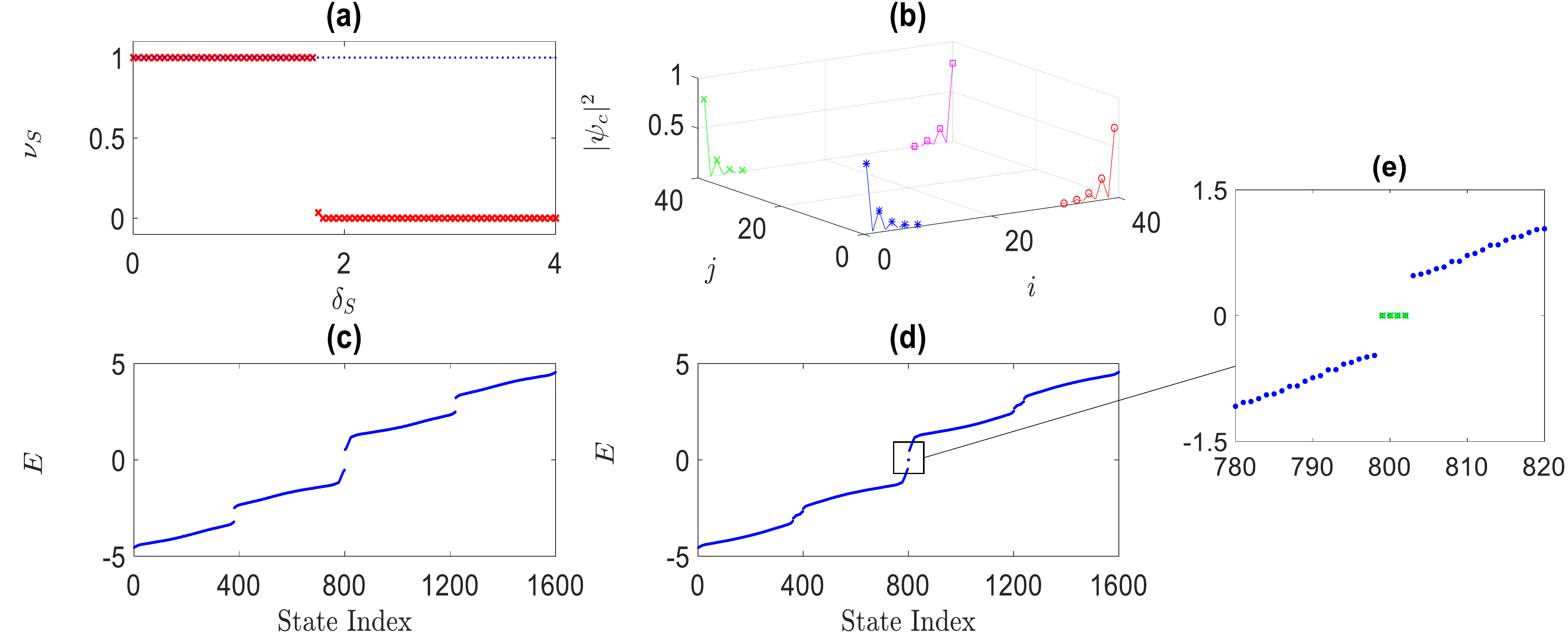}
	 	\end{center}
	 	\caption{(a) Topological invariant of $h_{S,k}$ as a function of $\delta_S$ {  (symmetry breaking perturbation strength)} under two different set of parameter values, where red crosses (blue dots) correspond to $J_S=0.75$ ($J_S=1.475$) and $\delta J_S=0.25$ ($\delta J_S=1.375$). (b) {  Probability distribution of each corner mode (marked with different colors) of $\mathcal{H}$ obtained in panel (d). For clarity, only $|\psi_c(i,j)|^2>0.005$ are shown.} (c) Energy level distribution of $\mathcal{H}$ when $h_{x,k}$ is topologically trivial, i.e., $J_x=0.75$, $J_y=1.475$, $\delta J_x=-0.25$, $\delta J_y=1.375$, $\delta_x=0.2$ and $\delta_y=0.15$. (d) Same as panel (c) but with $\delta J_x=0.25$, so that both $h_{x,k}$ and $h_{y,k}$ are topologically nontrivial. {  Panel (e) highlights the region near $E=0$ in panel (d), with corner modes highlighted in green.}}
	 	\label{pic} 
	 \end{figure*}

	\n where $H_{S,k}\equiv h_{a,S}+\mathrm{i} h_{b,S}$. It is well-known that in SSH model, the winding number $\nu_S=1$ ($\nu_S=0$) when the dimerization parameter $\delta J_S>0$ ($\delta J_S<0$). This again implies that corner states of $\mathcal{H}$ exist only if both dimerization parameters $\delta J_y,\delta J_x>0$. In order to check the generality of the above argument, we will now introduce a perturbation which amounts to modifying $h_{b,S}\rightarrow h_{b,S}'=h_{b,S}+\delta_S \cos k_S\sigma_y^{(S)}$. {  These extra terms break all but the chiral symmetry of the system.} Consequently, Eq.~(\ref{wn}) is still well-defined, which is plotted as a function of the perturbation strength $\delta_S$ in Fig.~\ref{pic}(a). In particular, we find that even for moderate perturbation strength, $h_{S,k}$ could still preserve its winding number. Consequently, by choosing different parameter values for $h_{x,k}$ and $h_{y,k}$, assuming both the presence of the perturbations $\delta_x$ and $\delta_y$, we find that $\mathcal{H}$ host topological corner modes only if $h_{x,k}$ and $h_{y,k}$ are both topologically nontrivial, i.e., $\nu_x=\nu_y=1$, as shown in Figs.~\ref{pic}(d) and (e). {  In general, $\delta_S$ can only induce topological transition if it is strong enough such that it closes the gap of $h(S,k)$. This happens once $\delta_S$ reaches $(J_S+\delta J_S)\sqrt{(J_S+\delta J_S)^2/(J_S-\delta J_S)^2-1}$, as can be verified from Fig.~\ref{pic}}(a).  
	
	 {  From the above discussion, the number of topological corner modes at zero-energy is then given by $n_0=4\nu$, { where $\nu=\nu_x\cdot \nu_y$ is a bulk invariant which accounts the \textit{bulk-corner correspondence} of our system.} If either or both $h_{x,k}$ and $h_{y,k}$ are topologically trivial, there is no such corner modes (see Fig.~\ref{pic}(c)). Finally, in Fig.~\ref{opic} we have also plotted the energy band structure of $\mathcal{H}$ under mixed boundary conditions, i.e., PBC along one direction and OBC along the other. Three representative cases have been considered in Fig.~\ref{opic}: (i) both directions are topologically trivial (panel (a) and (d)), (ii) only the $x$-direction is topologically nontrivial (panel (b) and (e)), and (iii) both directions are topologically trivial (panel (c) and (f)). As expected, both edge and bulk bands are gapped, and bulk or edge bands at zero energy are absent in all cases.

	\begin{figure}
		\begin{center}
			\includegraphics[scale=0.3]{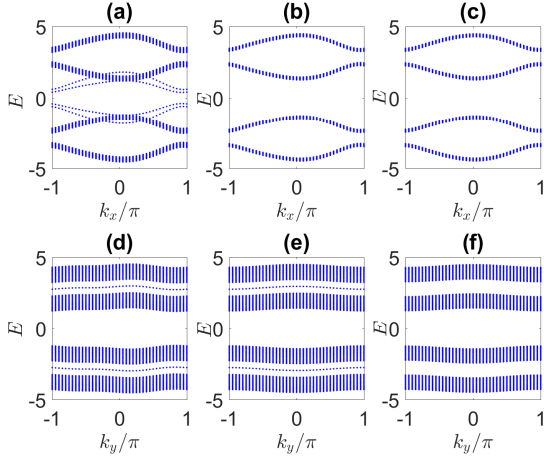}
		\end{center}
		\caption{(a,b,c) Energy band structure of Eq.~(\ref{SSH}) under OBC in the $y$-direction and PBC in the $x$-direction with (a) $\delta J_x=0.25$ and $\delta J_y=1.375$, (b) $\delta J_x=-0.25$ and $\delta J_y=1.375$, (c) $\delta J_x=-0.25$ and $\delta J_y=-1.375$. (d,e,f) Same as panel (a,b,c) but with OBC in the $x$-direction and PBC in the $y$-direction. Other parameters are the same in all panels with $J_x=0.75$, $J_y=1.475$, symmetry breaking perturbation strengths $\delta_x=\delta_y=0.1$, and Kronecker sum breaking perturbation strengths $\delta_{xy,1}=\delta_{xy,2}=0.1$.}
		\label{opic} 
	\end{figure}
	
}
	\subsection{Discussion} 
	\label{SSOTI3}
	
	In contrast to many existing proposals on SOTIs so far, our construction above introduces an SOTI model that is protected solely by the chiral symmetry and does not rely on any spatial symmetries. Therefore, our proposed model is fundamentally different from other SOTI models, such as those studied in Ref.~\cite{HTI1,HTI2,HTI3,HTI4,HTI5,HTI6,HTI7,HTI8,HTI9,HTI10,HTI11,HTI12,HTI13,HTI14,HTI15,HTI16,HTI17,HTI18,HTI19,HTI20,HTI21}, which belong to a family of second order topological crystalline insulators. In fact, our model closely resembles that of Ref.~\cite{LU}, which also relies on the existence of chiral symmetry alone. 
	
	While the model proposed in Ref.~\cite{LU} also describes a stack of 1D SSH models, it cannot be expressed as a Kronecker sum in the spirit of Eq.~(\ref{mstat}). However, since it can be broken down into four distinct 1D SSH models, the existence of corner modes is dictated by the four winding numbers of these SSH models. By contrast, the Kronecker sum structure of our model implies that only two winding numbers are needed to predict the existence of corner modes. Moreover, our model can be generalized beyond the Kronecker sum of two SSH models as described in Eq.~(\ref{mstat}). For example, as we will elucidate further in the next section, we may take $h_{x,k}$ to be a 1D Floquet topological insulator, which enables the construction of Floquet SOTIs.
	
	On the other hand, the nature of our construction as a stack of 1D topological phase to create a higher dimensional topological phase may at first seem reminiscent of weak topological insulators \cite{WTI}. This in turns raise an important question as to whether our construction merely represents a weak higher-order topological insulator. To answer this question, we first point out that edge states in typical weak topological insulators originate from a lower dimensional topological invariant, and as such are less robust against perturbations coupling a pair of their lower dimensional constituents. By contrast, the corner states in our model incorporate the interplay of two orthogonal coupling between a pair of 1D topological phase in the $x$- and $y$-directions, which leads to a bulk invariant defined in Sec.~\ref{SSOTI2} that depends on the topology in both directions. In particular, due to the topological protection in both $x$- and $y$-directions, such corner modes are robust against general perturbations. {  This is to be compared with the edge states appearing in our system, as illustrated in Fig.~\ref{opic}. There, the presence of edge states in the $x$-($y$-)direction depends only on the winding number in the $x$-($y$-)direction, and are thus sensitive to the direction in which the boundary is opened (see e.g. Figs.~\ref{opic}(b) and (e)). Therefore, while our construction indeed utilizes first-order weak topological insulators as its building block, the resulting second-order topological insulator does not inherit their weak topological effect.} 
	
	Finally, we also note that our proposed bulk invariant is different from the bulk quadrupole invariant introduced in Ref.~\cite{HTI1,HTI2}, since the latter is always zero in our model. {  Similar to the connection between the polarization (thence winding number) and the quantization of charge pumping \cite{FDHE3}, we argue that our bulk invariant is related to the 2D charge pumping usually proposed to probe the second Chern number in the context of 4D quantum Hall effect \cite{FDHE,FDHE2}. To this end, we may start by constructing a Hamiltonian
	
	\begin{eqnarray}
	H &=& (c_1+\cos k_x+\cos \phi_1) \sigma_x^{(x)} +\sin k_x \sigma_y^{(x)} +\sin \phi_1 \sigma_z^{(x)}  \nonumber \\
	&+& (c_2+\cos k_y+\cos \phi_2) \sigma_x^{(y)} +\sin k_y \sigma_y^{(y)} +\sin \phi_2 \sigma_z^{(y)} \;, \nonumber \\ \label{deform}
	\end{eqnarray}	
	
	\n where $\phi_1$ and $\phi_2$ are tunable parameters, $c_1$ and $c_2$ are constants. Note that when $\phi_1=0,\pi$ and $\phi_2=0,\pi$, Eq.~(\ref{deform}) reduces to the Hamiltonian Eq.~(\ref{mstat}) in its topologically trivial or nontrivial region depending on the values of $c_1$ and $c_2$. The second Chern number of Eq.~(\ref{deform}) is simply the product of the first Chern numbers of the first and second lines, which under appropriate gauge choices are related to the their respective polarization according to \cite{FDHE3}
	
	\begin{equation}
	C_2 = 4 [P_x(\pi)-P_x(0)]\times [P_y(\pi)-P_y(0)]\;, \label{c2}
	\end{equation} 
	
	\n where $P_S(\phi)=\int \frac{dk}{2\pi\mathrm{i}} \langle k_S,\phi | \partial_{k_S} |k_S,\phi\rangle $, $S=x,y$, and 
	
	\begin{equation}
	|k_x,k_y,\phi_1,\phi_2\rangle = |k_x,\phi_1\rangle \otimes |k_y,\phi_2\rangle
	\end{equation}
	
	\n is the eigenstate of $H$ associated with the band in which $C_2$ is evaluated. As detailed in Appendix~\ref{app2}, the polarization $P_S(\phi)$ at the chiral symmetric points, i.e., $\phi=0,\pi$, is related to the winding number $\nu_S(\phi)=2P_S(\phi)$. Equation~(\ref{c2}) can then be written as
	
	\begin{equation}
	C_2 = \nu(\pi,\pi)-\nu(0,\pi)-\nu(\pi,0)+\nu(0,0) \;, \label{ch2}
	\end{equation} 
	
	\n where $\nu(\phi_1,\phi_2)=\nu_x(\phi_1)\nu_y(\phi_2)$. 
	
	Since Eq.~(\ref{deform}) represents a Kronecker sum of two Chern insulating model, its second Chern number is readily obtained as \cite{FDHE3}
	
	\begin{equation}
	C_2=\begin{cases}
	1 & 2>c_1,c_2>0\text{ or }0>c_1,c_2>-2  \\
	0 & |c_1|>2\text{ or }|c_2|>2 \\
	-1 & 2>c_1,-c_2>0\text{ or }0>-c_1,c_2>-2
	\end{cases}\;.
	\end{equation}
	
	\n It can also be verified that when $|c_1|>2$ ($|c_2|>2$), both $\nu_x(0)$ and $\nu_x(\pi)$ ($\nu_y(0)$ and $\nu_y(\pi)$) are zero, so that $\nu(\pi,\pi)=\nu(0,0)=\nu(0,\pi)=\nu(\pi,0)=0$. If both $|c_1|<2$ and $|c_2|<2$, one of $\nu(\pi,\pi),\nu(0,0),\nu(0,\pi)$, or $\nu(\pi,0)$ is nonzero, while the other three are zero, so that $C_2$ is proportional to the nonvanishing bulk invariant $\nu$.
	
	Suppose now we modulate $\phi_2\rightarrow \phi_2 +Bx$, where $B$ is a constant, which simulates the presence of magnetic field perpendicular to $x$ and $\phi_2$ directions. By uniformly filling the lowest band of $H$, adiabatically tuning $\phi_1$ from $0$ to $2\pi$, the displacement of the particles in the $y$-direction, averaged over $\phi_2$ in $[0,2\pi]$, is proportional to $C_2$ \cite{FDHE}, and consequently also $\nu$ from the above argument. Our proposed bulk invariant thus provides a natural connection between the physics of higher-order topological phases and that of higher-dimensional topological phases. It will be interesting to explore the extension of this connection to a more general systems that do not enjoy Kronecker sum structure, which will be left for future work.
	}
	
	%{ (insert mean chiral displacement comments or results here)}
	
	We end this section by discussing the fate of our proposal in the presence of perturbations breaking the Kronecker sum structure of Eq.~(\ref{mstat}). To this end, we further add a perturbation of the form $h_{xy,k}=-\delta_{xy,1} \sigma^{(x)}_x \sigma^{(y)}_z-\delta_{xy,2}(\cos(k_x) \sigma^{(x)}_x +\sin(k_x) \sigma^{(x)}_y)\sigma^{(y)}_z$ to $\mathcal{H}$, which preserves its chiral symmetry but breaking the Kronecker sum structure of Eq.~(\ref{mstat}). In fact, we have also implemented such perturbations with $\delta_{xy,1}=\delta_{xy,2}=0.1$ in our results earlier presented in Figs.~\ref{pic}(b)-(e) and \ref{opic}. In general, we observe that the presence of small perturbations does not affect the existence of the topological corner modes in the system. At moderate perturbation strengths, however, it is possible for the bulk or edge band gaps to close, resulting in the change of the number of topological corner modes in the system, which can no longer be captured by the individual topological invariants of $h_x$ and $h_y$. Nevertheless, our results demonstrate that if a general 2D Hamiltonian can be adiabatically deformed into a Hamiltonian that admits Kronecker sum structure without closing the bulk or edge band gaps in the process, its higher-order topology can still be studied from the bulk perspective by calculating the topological invariants of two 1D Hamiltonian.  
	
	%\section{Constructing Floquet SOTI by stacking 1D Floquet topological phases}
	\section{Coupled-wire construction of Floquet SOTI} 
	\label{FloquetSOTI}
	
	\subsection{General formulation}
	\label{general}
	
	The idea we developed in Sec.~\ref{SSOTI} can also be applied to construct Floquet SOTIs. To this end, we may start with an array of chains of any 1D Floquet topological insulator in $x$-direction. Each of them is then coupled to adjacent chains by static dimerized couplings in $y$-direction~(see Fig.~\ref{stack} for an illustration). The full Hamiltonian of such a Floquet SOTI can then be written as
	\begin{eqnarray}
	\mathcal{H}(t) &=& - \sum_{j=1}^{N_y} \left\lbrace \sum_{i=1}^{N_x} \left[J_y + (-1)^j \delta J_y\right] |i,j+1\rangle \langle i,j | \right. \nonumber \\
	&& + \left. H_{1D}(t) \otimes |j\rangle \langle j | +\rm{h.c.} \right\rbrace \;, \label{fhoti1}
	\end{eqnarray}
	
	\n where $J_y\pm \delta J_y$ again denote the dimerized hopping amplitudes in the $y$-direction, and $H_{1D}(t)$ is a time-periodic Hamiltonian describing a 1D Floquet topological insulator. $\mathcal{H}(t)$ in Eq.~(\ref{fhoti1}) is thus time-periodic, and Floquet theory can be applied~\cite{Flo1,Flo2}. To this end, we define a Floquet operator as the one-period propagator
	\begin{equation}
	U_\mathcal{H} \equiv U(t+T,t) = \mathcal{T} \exp\left(-\mathrm{i} \int_t^{t+T} \mathcal{H}/\hbar  dt \right) \;,
	\end{equation}
	
	\n where $T$ is the period of the system in time, and $\mathcal{T}$ is the time-ordering operator. The topology of the system is then encoded in the quasienergy eigenstates $|\varepsilon\rangle$ of $U_\mathcal{H}$, which satisfies $U_\mathcal{H} |\varepsilon\rangle = e^{-\mathrm{i} \varepsilon T/\hbar} |\varepsilon\rangle$, where $\varepsilon$ is the associated quasienergy. 
	
	Since the first and second terms of Eq.~(\ref{fhoti1}) commute, we may write the Floquet operator as
	\begin{equation}
	U_\mathcal{H} = U_{H_{1D}}\otimes U_{H_y}\;, \label{fhoti2} 
	\end{equation} 
	
	\n where $U_{H_{1D}}$ and $U_{H_y}$ are 1D Floquet operators associated with $H_{1D}$ and $H_y=\sum_{j=1}^{N_y} \left\lbrace \left[J_y + (-1)^j \delta J_y\right] |j+1\rangle \langle j|+\rm{h.c.} \right\rbrace$, respectively. The tensor product structure of Eq.~(\ref{fhoti2}) enables us to systematically study the emergence of Floquet SOTIs from the properties of the underlying 1D Floquet system described by $H_{1D}$. Indeed, let $|0_{1(N_x)}\rangle $ and $|\pi_{1(N_x)}\rangle$ be the quasienergy zero and $\frac{\pi}{T}$ eigenstates of $U_{H_{1D}}$ localized near the left (right) end of the 1D lattice along $x$-direction. Topological corner modes of Eq.~(\ref{fhoti2}) at quasienergies zero and $\frac{\pi}{T}$ can then be obtained as {  (see also Appendix~\ref{app})}
	\begin{eqnarray}
	|0_{(X,Y)}\rangle &=& \sum_{j=1}^{N_y/2} (-1)^{j-1} \left(\frac{J_y-\delta J_y}{J_y+\delta J_y}\right)^{j-1} |0_X\rangle \otimes |Y_j\rangle \;, \nonumber \\
	|\pi_{(X,Y)} \rangle &=& \sum_{j=1}^{N_y/2} (-1)^{j-1} \left(\frac{J_y-\delta J_y}{J_y+\delta J_y}\right)^{j-1} |\pi_X\rangle \otimes |Y_j\rangle \;, \nonumber \\ \label{fcmode} 
	\end{eqnarray} 
	
	\n where $X=1,N_x$, $Y=1,N_y+2$, and $Y_j=|Y-2j|$. Equation~(\ref{fcmode}) thus shows that topological corner modes exist provided $|0_X\rangle$ and/or $|\pi_X\rangle$ exist and $\delta J_y>0$, i.e., \emph{both} $U_{H_{1D}}$ and $U_{H_y}$ are topologically nontrivial. 
	
	\subsection{Floquet SOTI with arbitrarily many topological corner modes}
	\label{FSOTI}
	
	To elucidate the application of our construction outlined in Sec.~\ref{general}, we consider a specific $H_{1D}(t)$ hereinafter as given by 
	\begin{eqnarray}
	H_{1D}(t) &=& \begin{cases}
h_1 & \text{ for } (m-1)T < t\leq (m-1/2)T \\
h_2 & \text{ for } (m-1/2)T < t\leq mT
\end{cases} \;, \nonumber \\
h_1 &=& J_1/ 2 \sum_{i,\sigma} \left[ |i, \sigma \rangle \langle i+1,\bar{\sigma} | + \rm{h.c.}\right] \;, \nonumber \\
h_2 &=& J_2/ (2\mathrm{i}) \sum_{i,\sigma} \left[ |i, \sigma \rangle \langle i+1,\bar{\sigma} | - \rm{h.c.}\right] \;, 
\label{H1D}
	\end{eqnarray}
	
	\n where $J_1$ and $J_2$ are hopping amplitudes, $\sigma=A,B$ is a sublattice or pseudospin index, $\bar{\sigma}$ is its complement, {  and $T$ is the period of the system, which will be taken as $T=2$ unless otherwise specified}. The model in Eq.~(\ref{H1D}) is first proposed in Ref.~\cite{LW} as a quantum-walk realization of spin-$1/2$ double kicked rotor, and later also extended to non-Hermitian~\cite{LW2} and 2D~\cite{LW3} systems. It is capable of hosting a controllable number of edge states. This can be shown by first writing down $h_1$ and $h_2$ in Eq.~(\ref{H1D}) in momentum space as
	\begin{eqnarray}
	h_{1,k} &=& J_1 \cos(k_x) \sigma^{(x)}_x \;, \nonumber \\
	h_{2,k} &=& J_2 \sin(k_x) \sigma^{(x)}_y \;.
\end{eqnarray}
	
	\n The momentum space Floquet operator of Hamiltonian Eq.~(\ref{H1D}) can then be found as~\cite{LW} (we take $\hbar=1$ from here onwards)
	\begin{equation}
	U_{H_{1D},k} =\exp\left(-\mathrm{i} h_{2,k}\right)\exp\left(-\mathrm{i} h_{1,k}\right) = \exp\left(-\mathrm{i} h_{\mathrm{eff},k}\right) \;, \label{mflo}
	\end{equation}
	
	\n where 
	\begin{equation}
	h_{\mathrm{eff},k} \propto \varepsilon = \arccos\left[\cos\left(J_1 \cos(k_x)\right)\cos\left(J_2 \sin(k_x)\right)\right] \;.\label{Dispersion}
	\end{equation}
	
	\n It follows that the quasienergy gap closes at $\varepsilon=0(\frac{\pi}{T})$ when $\cos\left(J_1 \cos(k_x)\right)\cos\left(J_2 \sin(k_x)\right)=+1(-1)$. Consequently, as one fixes $J_1$ ($J_2$), the two quasienergy gaps of $U_{H_{1D},k}$ close and reopen alternately at $\varepsilon=0$ and $\varepsilon=\frac{\pi}{T}$ when $J_2$ ($J_1$) increases by $\pi$. Every time the gap closes and reopens, a topological phase transition happens and new pairs of degenerate edge states at quasienergy zero or $\frac{\pi}{T}$ (i.e., Floquet zero or $\pi$ edge modes) emerge at both ends of the lattice~\cite{LW}. {  In particular, we find that each topological phase transition gives rise to two new pairs of either Floquet zero or $\pi$ modes. This can be understood from the fact that the eigenstates associated with each quasienergy band of $U_{H_{1D},k}$ always have a constant winding of $\pm 1$ despite the closing and reopening of the gap. By the bulk-edge correspondence, this eigenstate winding is associated with the difference between the number of Floquet edge modes at the two gaps in the quasienergy Brillouin zone. The emergence of two pairs of new edge modes at each topological phase transition is therefore necessary to preserve this eigenstate winding.}
	
\begin{figure*}
		\begin{center}
			\includegraphics[scale=0.6]{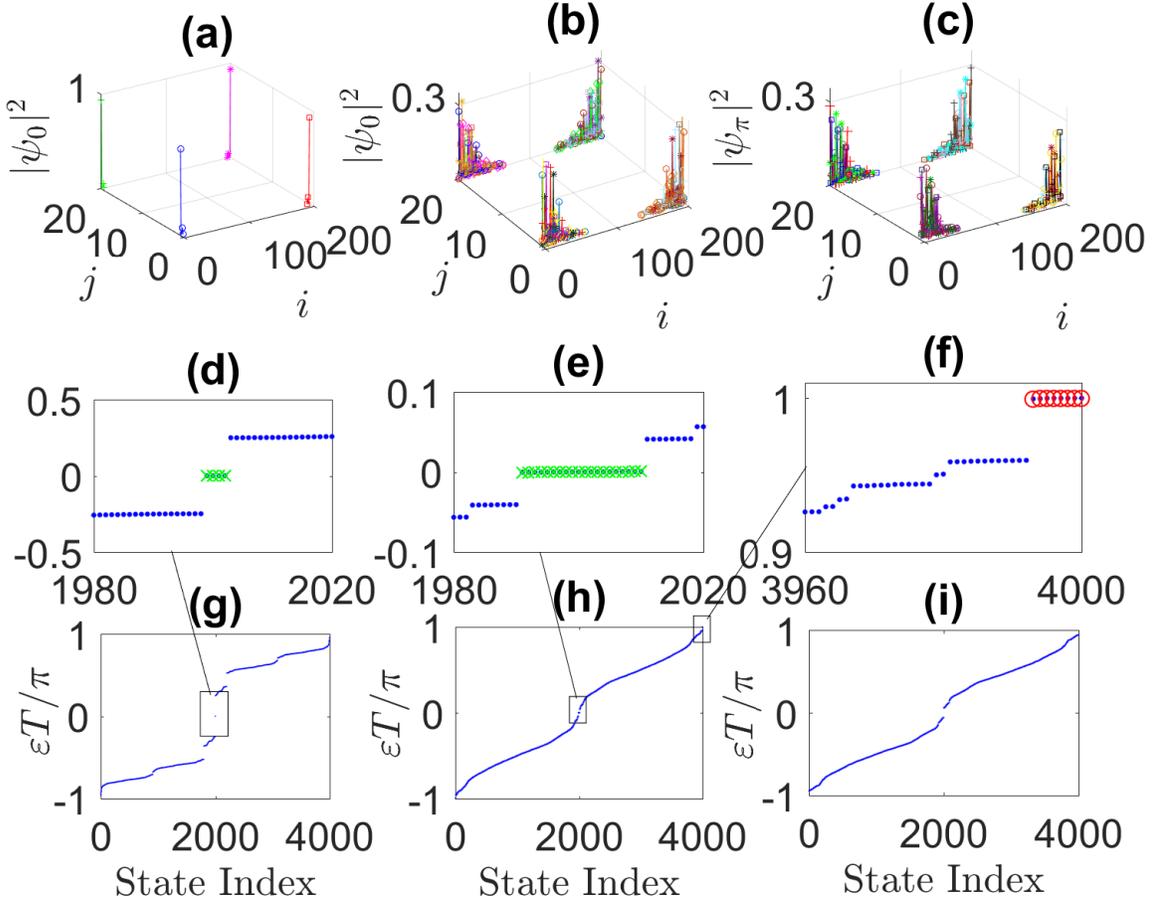}
		\end{center}
		\caption{   Panel (a) (panel (b)) shows the probability distribution of four (twenty) corner modes at quasienergy zero of $U_{\mathcal{H}}$ shown in panel (d) (panel (e)). Panel (f) shows the probability distribution of sixteen corner modes at quasienergy $\frac{\pi}{T}$ of $U_{\mathcal{H}}$ shown in panel (f). (g) Full quasienergy level distribution of $U_{\mathcal{H}}$ with $J_2=1$ and $\delta J_y=0.6875$, with corner modes highlighted in panel (d). (h) Full quasienergy level distribution of $U_{\mathcal{H}}$ with $J_2=14.8$ and $\delta J_y=0.6875$, with corner modes highlighted in panels (e) and (f). (i) Same as (h) but with $\delta J_y=-0.6875$, so that no corner modes is present. Other parameters are the same in all panels with $J_1=1$, $J_y=0.7375$, symmetry breaking perturbation strengths $\delta_x=\delta_y=0.1$, and Kronecker sum breaking perturbation strengths $\delta_{xy,1}=\delta_{xy,2}=0.1$. 
			}
			%Panel (a) and (b) ((c) and (d)) show the cumulative probability density $|\psi(i)|^2=\sum_n |\phi_n(i)|^2$ ($|\psi(i,j)|^2=\sum_n |\phi_n(i,j)|^2$) of all the edge (corner) states $\phi_n$ of $U_{\rm 1D}$ ($U_{\mathcal{H}}$) associated with Eq.~(\ref{H1D}) at quasienergy zero (blue) and $\frac{\pi}{T}$ (red). Panel (e)-(g) depict all the quasienergies of $U_{\mathcal{H}}$, with green circles and red crosses marking its Floquet zero and $\pi$ corner modes. The parameters chosen are $J_1=1$ and (a) $J_2=1$ (two zero modes), (b) $J_2=14.8$ (ten zero modes and eight $\pi$ modes), (c) and (e) $J_2=1$, $J_y=1.475$, and $\delta J_y=1.375$ (four zero corner modes), (d) and (f) $J_2=14.8$, $J_y=1.475$, and $\delta J_y=1.375$ (twenty zero corner modes and sixteen $\pi$ corner modes), (g) $J_2=14.8$, $J_y=1.475$, and $\delta J_y=-1.375$ (no corner modes).}
		\label{pic1} 
	\end{figure*}

	By implementing $H_{1D}(t)$ defined above to Eq.~(\ref{fhoti1}), the discussion of Sec.~\ref{general} implies the generation of Floquet SOTIs with arbitrarily many zero and $\pi$ corner modes satisfying Eq.~(\ref{fcmode}), whose number is also controllable via tuning the system parameters $\delta J_y$ and $J_1$ or $J_2$. {  This is also evidenced by our numerics as shown in Fig.~\ref{pic1}, in which the existence of zero and $\pi$ edge modes of $U_{H_{1D}}$ discussed earlier directly translates into a pair of the same number of Floquet zero and $\pi$ corner modes of $U_{\mathcal{H}}$ in panels (d)-(h), provided $\delta J_y>0$ (panel (h) vs. (i)).}
	
\begin{figure}
	\begin{center}
		\includegraphics[trim=1.7cm 7.6cm 1.2cm 8.5cm, clip=true, height=!,width=9cm]{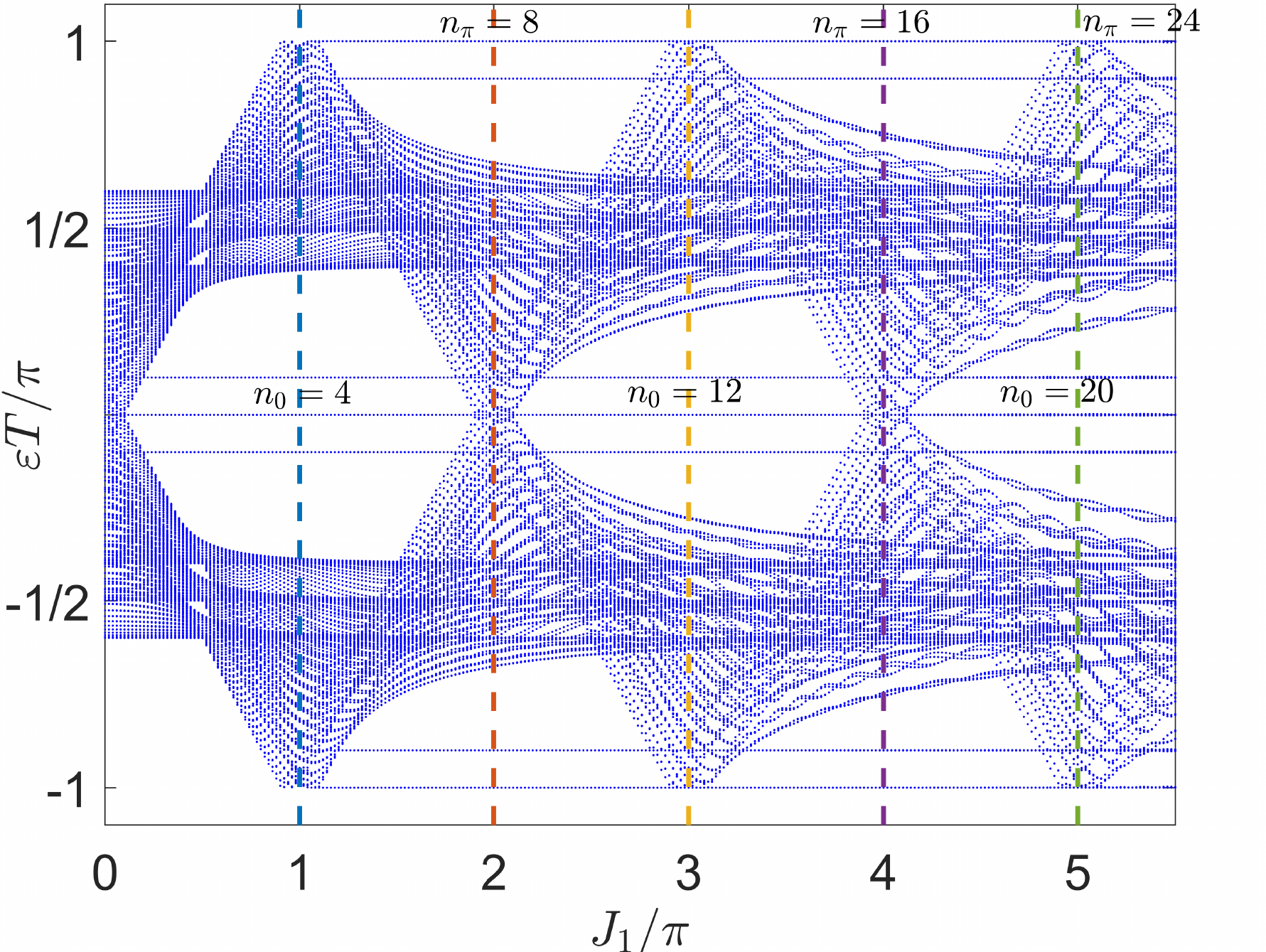}
	\end{center}
	\caption{The Floquet spectrum { $\varepsilon$} of $U_{\mathcal H}$ versus the hopping amplitude $J_1$. The size of the 2D lattice is $N_x=N_y=50$. Other system parameters are chosen as $J_y=\delta J_y=\pi/40$ and $J_2=\pi/2$. Dashed lines represent boundaries separating different Floquet SOTI phases in the parameter space. $n_0\;(n_\pi)$ denotes the number of Floquet topological corner states at quasienergy zero ($\frac{\pi}{T}$).}
	\label{Spectrum_vs_J1} 
	\end{figure}

	To further demonstrate the flexibility of generating many Floquet zero and $\pi$ corner modes following our construction, we show in Fig.~\ref{Spectrum_vs_J1} the quasienergy spectrum of $U_{\mathcal H}$ vs.~$J_1$, with the number of corner modes $n_0$ and $n_\pi$ denoted explicitly in the figure. Other system parameters are chosen as $J_y=\delta J_y=\pi/40$ and $J_2=\pi/2$. We find that with the increase of $J_1$, the number of Floquet corner modes $n_0\,(n_\pi)$ at quasienergy zero~($\frac{\pi}{T}$) increases by $8$ every time when $J_1$ passes through an even (odd) multiple of $\pi$. The same pattern is also observed in the Floquet spectrum of $U_{\mathcal H}$ vs.~$J_2$. {  This agrees with our discussion earlier that four new zero or $\pi$ edge modes emerge every time the quasienergy gap of $U_{H_{1D},k}$ closes and reopens, which translates into eight zero or $\pi$ corner modes when $U_{H_y}$ is topologically nontrivial. From the above discussion, we note that any number $n_0=4+8N$ of zero modes and $n_\pi=8N$ of $\pi$ modes can therefore be generated by simply setting either $(N+1)\pi<J_1<(N+2)\pi$ or $(N+1)\pi<J_2<(N+2)\pi$.} %{ In particular, we find that the number of Floquet corner modes at quasienergy zero ($\frac{\pi}{T}$) is given by $n_0=4|\nu_0\cdot \nu_y|\,(n_\pi=4|\nu_\pi \cdot \nu_y|)$, where $\nu_y$ is the winding number defined in Sec.~\ref{SSOTI2}, $\nu_0$ and $\nu_\pi$ are two Floquet winding numbers of $U_{H_{1D}}$ that will be introduced in the following subsection. At a fixed value of $J_2<\pi$ ($J_1<\pi$), $\nu_0$ and $\nu_\pi$ satisfy $(\nu_0-1)\pi<J_1<(\nu_0+1)\pi$ and $(\nu_\pi-1)\pi<J_1<(\nu_\pi+1)\pi$ ($(\nu_0-1)\pi<J_2<(\nu_0+1)\pi$ and $(\nu_\pi-1)\pi<J_2<(\nu_\pi+1)\pi$) respectively.}
	
\begin{figure}
	\begin{center}
		\includegraphics[scale=0.3]{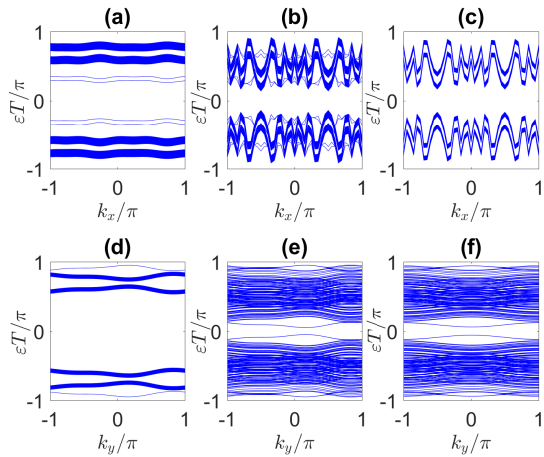}
	\end{center}
	\caption{(a,b,c) Quasienergy band structure of Eq.~(\ref{fhoti1}) under OBC in the $y$-direction and PBC in the $x$-direction with (a) $J_1=1$, $J_2=1$, and $\delta J_y=0.6875$ (b) $J_1=1$, $J_2=14.8$ and $\delta J_y=0.6875$, (c) $J_1=1$, $J_2=14.8$ and $\delta J_y=-0.6875$. (d,e,f) Same as panel (a,b,c) but with OBC in the $x$-direction and PBC in the $y$-direction. Other parameters are the same in all panels with $J_y=0.7375$, symmetry breaking perturbation strengths $\delta_x=\delta_y=0.1$, and Kronecker sum breaking perturbation strengths $\delta_{xy,1}=\delta_{xy,2}=0.1$.}
	\label{opic2} 
\end{figure}

	Note in passing that edge states are also found at quasienergies $\pm2(J_y+\delta J_y)$ and $\pm(\pi-2(J_y+\delta J_y))$ in Fig.~\ref{Spectrum_vs_J1}, which are visible whenever OBC is applied in the $x$-direction. These edge states are remnants of the Floquet zero and $\pi$ edge states of $U_{H_{1D}}$, which are shifted from quasienergy zero or $\frac{\pi}{T}$ by the coupling in the $y$-direction. As a result, they lose their topological protection and are distinguished from the Floquet zero and $\pi$ corner modes which remain pinned at quasienergy zero and $\pi$. {  The behavior of these edge states under a more general choice of parameters can be inferred from the quasienergy spectrum under mixed boundary conditions, i.e., PBC in one direction and OBC in the other, shown in Fig.~\ref{opic2}. In particular, it can be observed that with $J_y\neq \delta J_y$, the quasienergy gaps when OBC is applied along the $x$ direction become smaller with the increase in the number of zero and $\pi$ corner modes. This can be understood from the fact that the many Floquet zero and $\pi$ edge states of $U_{H_{1D}}$ are shifted differently from their respective quasienergy zero or $\frac{\pi}{T}$. This results in many edge states filling in the finite size of the quasienergy Brillouin zone, which consequently leads to smaller quasienergy gaps, as seen from Figs.~\ref{opic2}(e) and (f). Although the quasienergy gaps about $\varepsilon=0$ and $\varepsilon=\pi/T$ are not visible in these two panels, we have calculated that they are still finite, i.e., $\Delta \varepsilon_{0,{\pi/T}}\approx 0.15$. This justifies the term Floquet SOTI for our model, since its topologically protected states are localized at the boundary of a boundary (the corner) and are separated from the edge and bulk bands by finite gaps.}%For instance, the  quasienergy gaps at the middle and edge of the quasienergy Brillouin zone are both found to be $\Delta \varepsilon_{0,\frac{\pi}{T}}\approx 0.15$ in Fig.~\ref{opic2}(h). } %This justifies the term Floquet SOTI for our model, since its topologically protected states are localized at the boundary of a boundary (the corner).  However, these edge states are gapped and their numbers depend on the size of the lattice. Therefore, they do not belong to the type of Floquet topological corner states that is the focus of our study.}
	
	\subsection{Symmetry analysis and topological invariant}  
	\label{flsoti}
	
	{In this subsection, we introduce the topological invariants characterizing the Floquet SOTIs, and discuss their relations to the number of Floquet corner states.} By transforming Eq.~(\ref{mflo}) to symmetric time frames~\cite{cref},
	\begin{eqnarray}
	\tilde{U}^{(1)}_{H_{1D},k} &=& \hat{F}_k \hat{G}_k\;, \tilde{U}^{(2)}_{H_{1D},k} = \hat{G}_k \hat{F}_k\;, \nonumber \\
	\hat{F}_k &=& \exp\left(-\mathrm{i} h_{2,k}/2\right)\times \exp\left(-\mathrm{i} h_{1,k}/2\right)  \;, \nonumber \\
	\hat{G}_k &=& \exp\left(-\mathrm{i} h_{1,k}/2\right)\times \exp\left(-\mathrm{i} h_{2,k}/2\right) \;,
	\end{eqnarray}
	
	\n The full 2D momentum space Floquet operator can be written as,
	\begin{equation}
	\tilde{U}^{(1,2)}_{\mathcal{H},k} = \tilde{U}^{(1,2)}_{H_{1D},k} \otimes \exp\left(-2\mathrm{i} h_{y,k}\right)  \;,
	\end{equation}
	
	\n where $h_{y,k}$ is defined in Eq.~(\ref{mstat}). In particular, it is easy to verify that both $\tilde{U}^{(1)}_{H_{1D},k}$ and $\tilde{U}^{(2)}_{H_{1D},k}$ possess inversion, time-reversal, particle-hole, and chiral symmetries, respectively, given by the same operators defined in Sec.~\ref{SSOTI2}, which satisfy $\mathcal{I}_x \tilde{h}_{1D}(k_x,t) \mathcal{I}_x^{-1}=\tilde{h}_{1D}(-k_x,t)$, $\mathcal{T}_x \tilde{h}_{1D}(k_x, t)\mathcal{T}_x^{-1}=\tilde{h}_{1D}(-k_x, 2-t)$,  $\mathcal{P}_x \tilde{U}_{H_{1D},k} \mathcal{P}_x^{-1}=\tilde{U}_{H_{1D},-k}$, and $\Gamma_x \hat{F}_k \Gamma_x= \hat{G}_k$ \cite{cref,DG,RG3,LW,FPT}, where $\tilde{h}_{1D}(k_x,t)$ is the momentum space time-dependent Hamiltonian in the symmetric time frame associated with Eq.~(\ref{H1D}). This implies that $\tilde{U}^{(1,2)}_{H_{1D},k}$ also belong to class BDI in the AZ classification scheme, which is now characterized by two winding numbers $\nu_0$ and $\nu_\pi$ associated with the number of Floquet zero and $\pi$ edge modes respectively \cite{FPT}. By writing $\tilde{U}^{(1)}_{H_{1D},k}$ explicitly as a matrix in the $\sigma^{(x)}_z$ basis,
	\begin{equation}
	\tilde{U}^{(1)}_{H_{1D},k}\hat{=} \left(\begin{array}{cc}
	a(k_x) & b(k_x) \\
	c(k_x) & d(k_x) \\
	\end{array}\right) \;,
	\end{equation}   
	
	\n the winding numbers $\nu_0$ and $\nu_\pi$ can be obtained as~\cite{cref,RG6}
	\begin{eqnarray}
	\nu_0 &=& \frac{1}{2\pi\mathrm{i}} \int dk_x b^{-1} \frac{d}{dk_x} b \;, \nonumber \\
	\nu_\pi &=& \frac{1}{2\pi\mathrm{i}} \int dk_x d^{-1} \frac{d}{dk_x} d \;.
	\end{eqnarray}
	
\begin{figure}
		\begin{center}
			\includegraphics[scale=0.3]{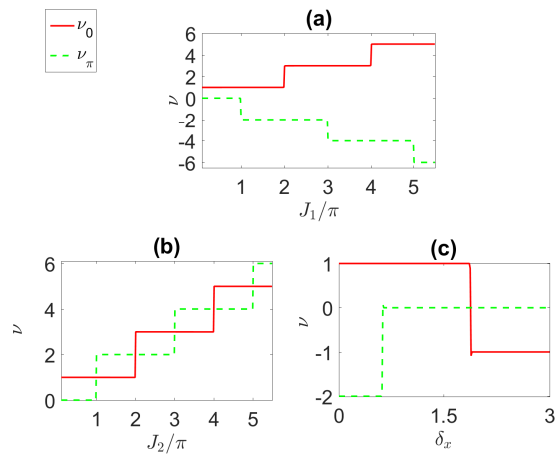}
		\end{center}
		\caption{Winding numbers $\nu_0$ (red solid lines) and $\nu_\pi$ (green dashed lines) as a function of (a) $J_1$, (b) $J_2$, (c) $\delta_x$.}
		\label{pic2} 
	\end{figure}	
	
	We plot $\nu_0$ and $\nu_\pi$ of $\tilde{U}^{(1)}_{H_{1D},k}$ as a function of $J_1$ ($J_2$) in Fig.~\ref{pic2}(a) (Fig.~\ref{pic2}(b)), where $J_2=1$ ($J_1=1$) is fixed. Consistent with the argument presented in Sec.~\ref{FSOTI}, either $\nu_0$ or $\nu_\pi$ increases as $J_1$ or $J_2$ increases by an integer multiple of $\pi$. In the presence of perturbation $h_{2,k}\rightarrow h_{2,k}-\delta_x \cos(k_x) \sigma_y^{(x)}$, which breaks all but chiral symmetry of the system, the winding numbers $\nu_0$ and $\nu_\pi$ remain well-defined, as depicted in Fig.~\ref{pic2}(c) at $J_1=5$ and $J_2=1$. A large number of zero and $\pi$ edge modes can therefore be generated in a controlled manner by simply setting $J_1$ or $J_2$ to be large~\cite{LW}. { From Eq.~(\ref{fcmode}), this implies the generation of arbitrarily many Floquet zero and $\pi$ corner modes, tunable via the parameters $\delta J_y$, $J_1$, and/or $J_2$. Recalling that $\nu_y$ is the winding number of $h_{y,k}$, bulk topological invariants of $\mathcal{U}_{\mathcal{H}}$ can then be constructed as $\nu_{0,y}=\nu_0\cdot \nu_y$ and $\nu_{\pi,y}=\nu_\pi \cdot \nu_y$, which determine the number of Floquet zero and $\pi$ corner modes as $n_0=4|\nu_{0,y}|$ and $n_\pi=4|\nu_{\pi,y}|$ respectively, thereby establishing the ``bulk-corner correspondence'' of our system.}
	
	In Figs.~\ref{pic1}(c)-(g), we have also included the presence of small perturbations of the form $h_{1,k}\rightarrow h_{1,k}-\delta_{xy,1} \cos(k_x) \sigma^{(x)}_x\sigma_z^{(y)}$, $h_{2,k}\rightarrow h_{2,k}-\delta_{xy,2} \sin(k_x) \sigma^{(x)}_y\sigma_z^{(y)}-\delta_x \cos(k_x) \sigma_y^{(x)}$, and $h_{y,k}\rightarrow h_{y,k}-\delta_y \cos(k_y) \sigma_y^{(y)}$, where $\delta_x$ and $\delta_y$ terms break all but chiral symmetry of the system, while $\delta_{xy,1}$ and $\delta_{xy,2}$ terms break the tensor product structure of Eq.~(\ref{fhoti2}). As expected, such perturbations do not qualitatively affect the existence of zero and $\pi$ corner modes in the system, provided the former do not induce edge or bulk gap closing of the quasienergy bands. This shows that our Floquet SOTI proposal does not rely on any spatial symmetry protection and its topological characterization presented above also provide insights into more general Floquet SOTI models, whose Floquet operator can be adiabatically deformed to the form of Eq.~(\ref{fhoti2}).

{ 	 

\section{Detection of bulk invariants} \label{detect}

As discussed in Sec.~\ref{SSOTI3}, our proposed bulk invariant $\nu$ in the static system manifests itself as the amount of charge displaced when the system is subject to adiabatic variations of two parameters over a cycle. Such a 2D charge pump has already been realized in photonic \cite{FDHE2} and cold-atom \cite{FDHE} setups for the study of 4D quantum Hall effect. An appropriate modification to these experiments is thus expected to be feasible for detecting $\nu$.  

Another promising means to detect the bulk topological invariants of Floquet SOTIs introduced in Sec.~\ref{flsoti} is to measure the mean chiral displacement (MCD) of a wavepacket, which will be detailed in the following.}

\subsection{Mean chiral displacement}\label{MCD}

The MCD is proposed in~\cite{MCD2} and applied in~\cite{LW,MCD1,MCD3} as a dynamical probe of winding
numbers for 1D topological insulators. The tensor product structure
of Floquet operator $U_{{\cal H}}$, together with its chiral symmetry
allow us to extend the definition of MCD straightforwardly to the two-dimensional
dynamics of our model.

We first introduce the chiral displacement operator $\hat{C}_{\alpha}$,
which in Heisenberg representation is given by 
\begin{equation}
\hat{C}_{\alpha}(t)=\left[U_{{\cal H}}^{(\alpha)}\right]^{-t}(\hat{x}\otimes\Gamma_{x})\otimes(\hat{y}\otimes\Gamma_{y})\left[U_{{\cal H}}^{(\alpha)}\right]^{t}.
\label{CDO}
\end{equation}
Here $U_{{\cal H}}^{(\alpha)}$ is the full Floquet operator given
by Eq.~(\ref{fhoti2}) in the symmetric time frame $\alpha$, $t$ denotes the
number of driving periods, $\hat{x}$ and $\hat{y}$ are quantized
unit-cell position operators. For the model we investigated in Sec.~\ref{FloquetSOTI}, the chiral symmetry operators $\Gamma_{x}$ and $\Gamma_{y}$
are explicitly given by $\Gamma_{x}=\sigma_{z}^{(x)}$ and $\Gamma_{y}=\sigma_{z}^{(y)}$. For a wavepacket $|\psi_{0}\rangle$ prepared at time
$t=0$, the expectation value $\langle\psi_{0}|\hat{C}_{\alpha}(t)|\psi_{0}\rangle$
thus describes the chirality-resolved shift of $|\psi_{0}\rangle$
over $t$s driving periods.

We now choose the initial state $|\psi_{0}\rangle$ to be a fully
polarized state located at the center ($x=0,y=0$) of the lattice \cite{note2}.
Explicitly it has the form
\begin{equation}
|\psi_{0}\rangle=|0_{x}\rangle\otimes|\pm_{x}\rangle\otimes|0_{y}\rangle\otimes|\pm_{y}\rangle,
\end{equation}
where $|0_{x}\rangle$ ($|0_{y}\rangle$) is the eigenstate of $\hat{x}$
($\hat{y}$) with eigenvalue $0$, and $|\pm_{x}\rangle$ ($|\pm_{y}\rangle$)
is the eigenstate of $\Gamma_{x}$ ($\Gamma_{y}$) with eigenvalue
$+1$ or $-1$. The MCD of such a wavepacket over $t$s driving periods
is then given by:
\begin{alignat}{1}
C_{\alpha}(t)= & \langle0_{x}|\otimes\langle\pm_{x}|\otimes\langle0_{y}|\otimes\langle\pm_{y}|\nonumber \\
\times & \left[U_{{\cal H}}^{(\alpha)}\right]^{-t}(\hat{x}\otimes\Gamma_{x})\otimes(\hat{y}\otimes\Gamma_{y})\left[U_{{\cal H}}^{(\alpha)}\right]^{t}\nonumber \\
\times & |0_{x}\rangle\otimes|\pm_{x}\rangle\otimes|0_{y}\rangle\otimes|\pm_{y}\rangle
\label{CAlphat}
\end{alignat}

To proceed, we express $U_{{\cal H}}^{(\alpha)}$ as 
\begin{equation}
U_{{\cal H}}^{(\alpha)}=U_{x}^{(\alpha)}\otimes U_{y}^{(\alpha)},
\label{UHAlpha}
\end{equation}
where $U_{x}^{(\alpha)}$ and $U_{y}^{(\alpha)}$ are 1D Floquet operators
associated with Hamiltonians $H_{1D}$ and $H_{y}$ in Eq.~(\ref{fhoti2}), respectively.
Note that for the time-independent Hamiltonian $H_{y}$, we have $U_{y}^{(1)}=U_{y}^{(2)}$.
With the help of Eq.~(\ref{UHAlpha}), we can rewrite $C_{\alpha}(t)$ as a product
of two MCDs along two orthogonal dimensions, i.e.,
\begin{equation}
C_{\alpha}(t)=C_{\alpha x}(t)\cdot C_{\alpha y}(t),
\label{CAlphatDCP}
\end{equation}
where 
\begin{alignat}{1}
C_{\alpha x}(t)= & \langle0_{x}|\otimes\langle\pm_{x}|[U_{x}^{(\alpha)}]^{-t}(\hat{x}\otimes\Gamma_{x})[U_{x}^{(\alpha)}]^{t}|0_{x}\rangle\otimes|\pm_{x}\rangle,\\
C_{\alpha y}(t)= & \langle0_{y}|\otimes\langle\pm_{y}|[U_{y}^{(\alpha)}]^{-t}(\hat{y}\otimes\Gamma_{y})[U_{y}^{(\alpha)}]^{t}|0_{y}\rangle\otimes|\pm_{y}\rangle.
\end{alignat}
Now performing a Fourier transform from position to momentum representations,
we find
\begin{alignat}{1}
C_{\alpha x}(t)= & \int_{-\pi}^{\pi}\frac{dk_{x}}{2\pi}\langle\pm_{x}|[U_{k_{x}}^{(\alpha)}]^{-t}\Gamma_{x}i\partial_{k_{x}}[U_{k_{x}}^{(\alpha)}]^{t}|\pm_{x}\rangle,\\
C_{\alpha y}(t)= & \int_{-\pi}^{\pi}\frac{dk_{y}}{2\pi}\langle\pm_{y}|[U_{k_{y}}^{(\alpha)}]^{-t}\Gamma_{y}i\partial_{k_{y}}[U_{k_{y}}^{(\alpha)}]^{t}|\pm_{y}\rangle,
\end{alignat}
where $U_{k_{x}}^{(\alpha)}$ and $U_{k_{y}}^{(\alpha)}$ are $2\times2$
matrices satisfying $U_{x}^{(\alpha)}=\sum_{k_{x}}U_{k_{x}}^{(\alpha)}|k_{x}\rangle\langle k_{x}|$
and $U_{y}^{(\alpha)}=\sum_{k_{y}}U_{k_{y}}^{(\alpha)}|k_{y}\rangle\langle k_{y}|$
in momentum representations. Then following the derivations detailed
in Ref.~\cite{LW}, we obtain
\begin{alignat}{1}
C_{\alpha x}(t)= & \frac{v_{\alpha}}{2}-\int_{-\pi}^{\pi}\frac{dk_{x}}{2\pi}\frac{\cos(2\varepsilon t)}{2}(n_{x}^{\alpha}\partial_{k_{x}}n_{y}^{\alpha}-n_{y}^{\alpha}\partial_{k_{x}}n_{x}^{\alpha}),\label{CAlphaxt}\\
C_{\alpha y}(t)= & \frac{w_{\alpha}}{2}-\int_{-\pi}^{\pi}\frac{dk_{y}}{2\pi}\frac{\cos(2Et)}{2}(d_{x}^{\alpha}\partial_{k_{y}}d_{y}^{\alpha}-d_{y}^{\alpha}\partial_{k_{y}}d_{x}^{\alpha}).\label{CAlphayt}
\end{alignat}
Here, for the Floquet model we studied in the last section, $v_{\alpha}$
is the winding number of the 1D Floquet operator $\tilde{U}_{H_{1{\rm D},k}}^{(\alpha)}$
in symmetric time frame $\alpha$, $w_{\alpha}=\nu_y$ is the winding
number of SSH model associated with the propagator $e^{-i2h_{y,k}}$,
$\varepsilon$ is the eigenphase of $\tilde{U}_{H_{1{\rm D},k}}^{(\alpha)}$
as defined in Eq.~(\ref{Dispersion}), and $E=\pm4J_{y}$ is the eigenphase
of $e^{-i2h_{y,k}}$. The components of unit vectors $(d_{x}^{\alpha},d_{y}^{\alpha})$
and $(n_{x}^{\alpha},n_{y}^{\alpha})$ are given by

\begin{equation}
d_{x}^{\alpha}=\cos k_{y}\qquad d_{y}^{\alpha}=\sin k_{y},
\end{equation}
and
\begin{alignat}{1}
n_{x}^{1} & =\frac{\sin({\cal J}_{1})}{\sin(\varepsilon)},\quad n_{y}^{1} =\frac{\sin({\cal J}_{2})\cos({\cal J}_{1})}{\sin(\varepsilon)},\nonumber \\
n_{x}^{2} & =\frac{\sin({\cal J}_{1})\cos({\cal J}_{2})}{\sin(\varepsilon)},\quad n_{y}^{2} =\frac{\sin({\cal J}_{2})}{\sin(\varepsilon)},
\end{alignat}
where ${\cal J}_{1}=J_{1}\cos k_{x}$ and ${\cal J}_{2}=J_{2}\sin k_{x}$.
It is clear that both $C_{\alpha x}(t)$ and $C_{\alpha y}(t)$ are
composed of a time-independent topological part and a time-dependent
oscillating term. For general dispersion relations, the oscillating
terms tend to decay at large $t$ under the integral over corresponding
quasimomentum.

To relate $C_{\alpha}(t)$ to the topological invariants of Floquet SOTIs, we consider its average over $t$s
driving periods, given by
\begin{equation}
\overline{C_{\alpha}(t)}=\frac{1}{t}\sum_{t'=1}^{t}C_{\alpha}(t').
\end{equation}
With the help of Eqs.~(\ref{CAlphatDCP}), (\ref{CAlphaxt}) and (\ref{CAlphayt}), we see that the oscillating
parts of $\overline{C_{\alpha}(t)}$ decay in time at least of order
$\frac{1}{t}$. Therefore, in long time limit ($t\rightarrow\infty$),
we obtain
\begin{equation}
\overline{C_{\alpha}}\equiv\lim_{t\rightarrow\infty}\overline{C_{\alpha}(t)}=\frac{v_{\alpha}w_{\alpha}}{4}.\label{CBarAlpha}
\end{equation}

For the model we considered in the last section, the winding number
$w_{\alpha}=w=1$. Furthermore, the winding numbers $v_{1}$ and $v_{2}$
are related to $\nu_{0}$ and $\nu_{\pi}$~\cite{cref} through
\begin{equation}
\nu_{0}=\frac{v_{2}+v_{1}}{2},\qquad\nu_{\pi}=\frac{v_{2}-v_{1}}{2}.\label{nu0pi}
\end{equation}
Combining Eqs.~(\ref{CBarAlpha}) and (\ref{nu0pi}) then yields the relations between time-averaged MCDs and topological winding numbers $\nu_{0,\pi}$, i.e.,
\begin{alignat}{1}
\nu_{0}= & 2(\overline{C_{2}}+\overline{C_{1}}),\nonumber\\
\nu_{\pi}= & 2(\overline{C_{2}}-\overline{C_{1}}).\label{nu-MCD}
\end{alignat}

Therefore, by measuring the long-time averaged MCDs in two  complementary
symmetric time frames, we would be able to obtain the topological
invariants characterizing the Floquet SOTIs introduced in Sec.~\ref{flsoti} \cite{MCD2}. The number of Floquet corner states can also be indirectly
deduced from bulk dynamics through the relations
\begin{alignat}{1}
n_{0}= & 8|\overline{C_{1}}+\overline{C_{2}}|,\\
n_{\pi}= & 8|\overline{C_{1}}-\overline{C_{2}}|.
\end{alignat}

It is also not hard to extend these results to other Floquet SOTIs
protected by chiral symmetry, for which evolutions in four
symmetric time frames may need to be executed. The formalism presented here
could also be applied to static SOTIs protected by chiral symmetry, where the
number of driving periods $t$ should be interpreted as the duration of evolution time, and the sum over $t$ replaced by an integral over the continuous time duration $t$.

\begin{figure}
	\begin{center}
		\includegraphics[scale=0.48]{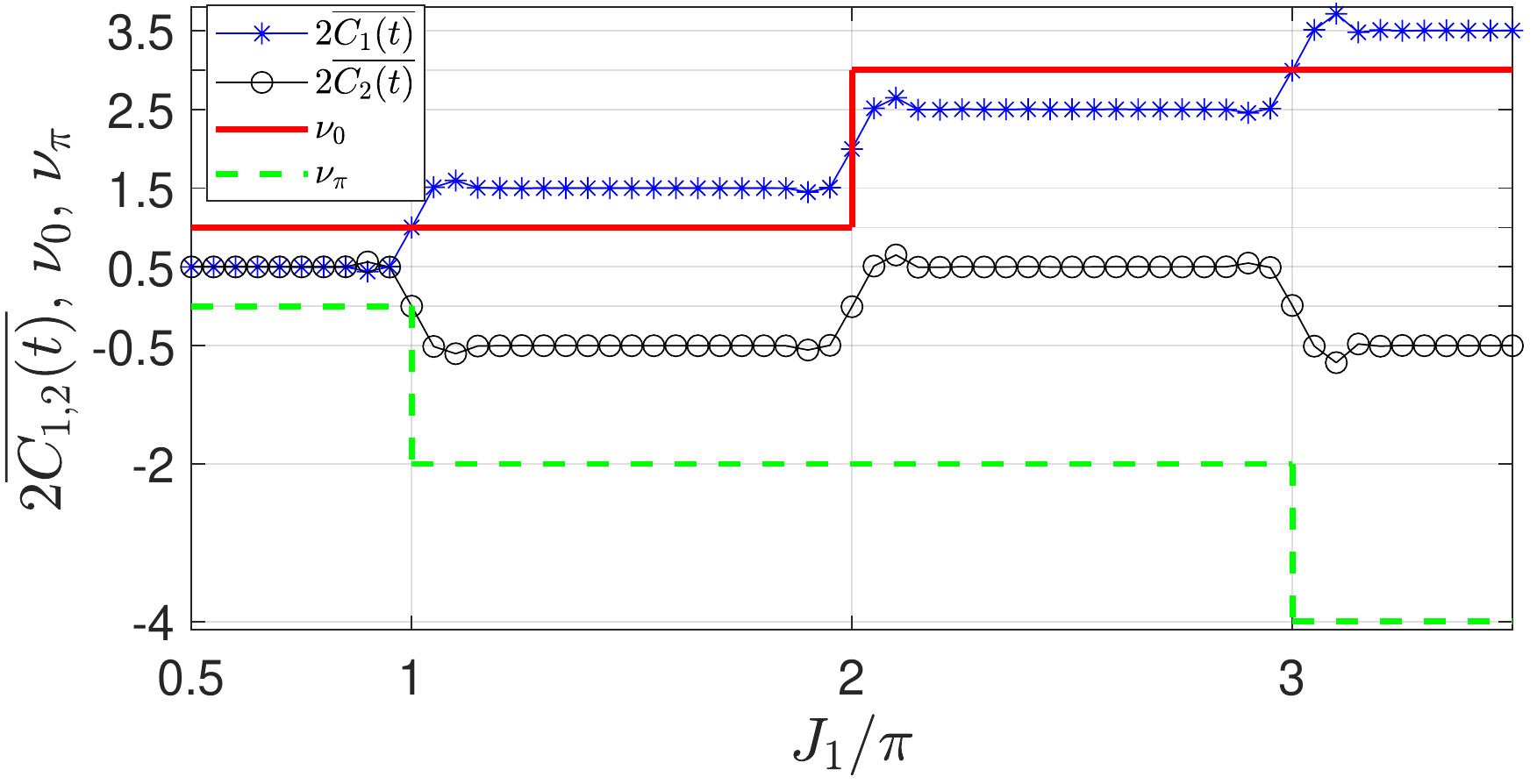}
	\end{center}
	\caption{MCDs and winding numbers versus the hopping amplitude $J_1$. Numerical results of $2\overline{C_1(t)}$ and $2\overline{C_2(t)}$, both averaged over $t = 20$ driving periods, are shown by the blue stars and black circles. Theoretical values of winding numbers $\nu_{0}$ and $\nu_{\pi}$ are denoted by the red solid and green dashed lines, respectively. Other system parameters are chosen as $J_y=\delta J_y=\pi/40$ and $J_2=\pi/2$. Topological phase transitions happen at $J_1=\pi,2\pi,3\pi$, away from which the relations in Eqs.~(\ref{nu-MCD}) are verified.}
	\label{figMCD} 
\end{figure}	

\begin{figure}
	\begin{center}
		\includegraphics[scale=0.48]{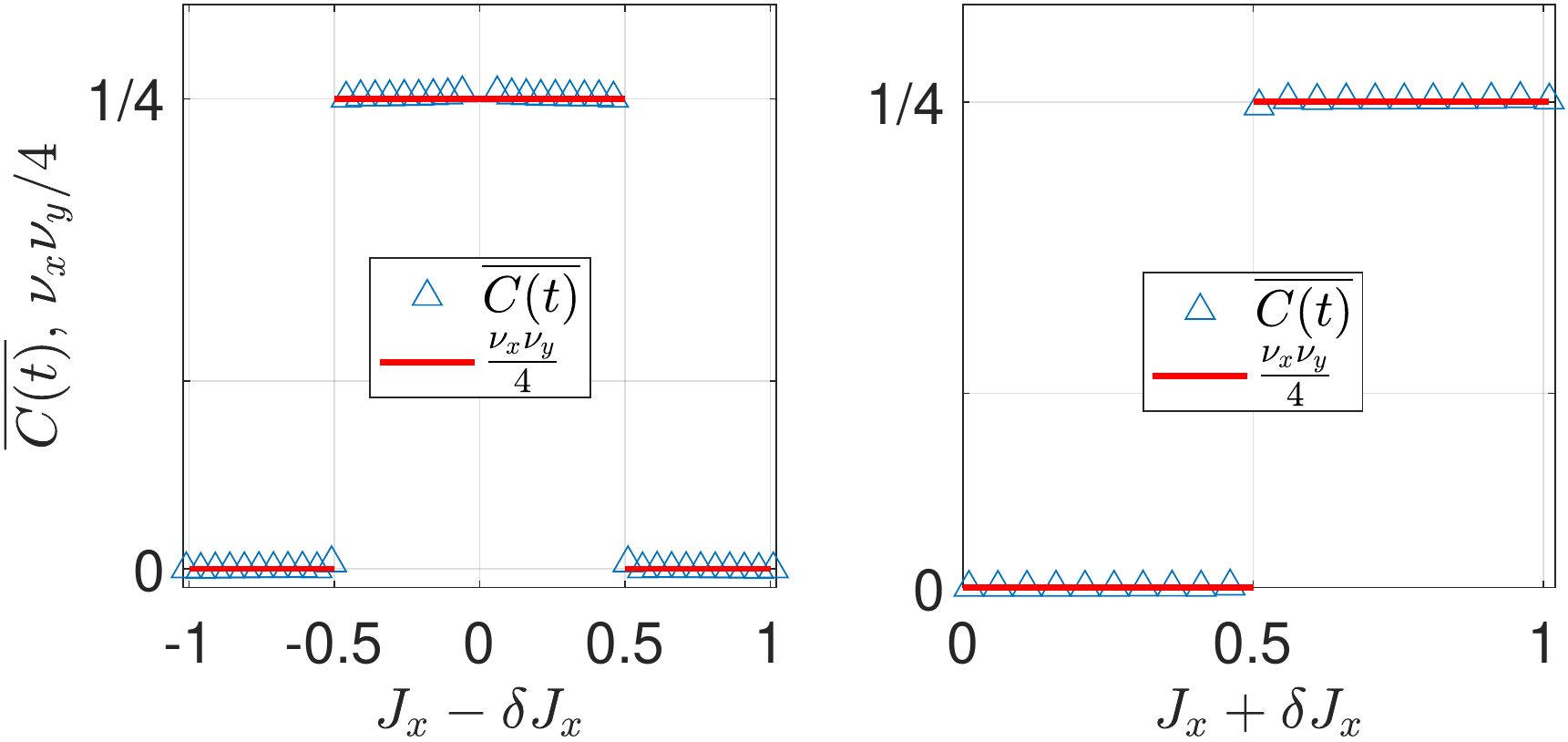}
	\end{center}
	\caption{MCDs of the static SOTI model~[defined in Eq.~(\ref{mstat})] averaged over a long time duration $\tau=400$ (triangles). Solid lines are theoretical values of $\nu_x\nu_y/4$, with the two winding numbers $\nu_x,\nu_y$ defined in Eq.~(\ref{wn}). Other system parameters are chosen as $J_x+\delta J_x=1/2,J_y-\delta J_y=0,J_y+\delta J_y=1/2$ in panel (a), and $J_x-\delta J_x=1/2,J_y-\delta J_y=0,J_y+\delta J_y=1/2$ in panel (b).}
	\label{figMCD2} 
\end{figure}	

In Fig.~\ref{figMCD}, we present the $\overline{C_{\alpha}(t)}$ and the
winding numbers $\nu_{0},\nu_{\pi}$ vs.~$J_{1}$. 
Up to $t=20$, we already find good convergence of $2\overline{C_{2}(t)}+2\overline{C_{1}(t)}$
and $2\overline{C_{2}(t)}-2\overline{C_{1}(t)}$ to their corresponding
winding numbers $\nu_{0}$ and $\nu_{\pi}$,
respectively. When $J_1$ is close to an integer multiple of $\pi$, the MCD combinations $2\overline{C_{2}(t)}+2\overline{C_{1}(t)}$ and $2\overline{C_{2}(t)}-2\overline{C_{1}(t)}$ deviate from quantization due to the topological phase transitions happening there. 
 {Other small deviations from quantization are finite time effects, which are originated from the oscillating terms
in Eqs.~(\ref{CAlphaxt}) and (\ref{CAlphayt}).
With the increase of $t$, the parameter windows around $J_1/\pi=1,2,3$ in which $2\overline{C_{1,2}(t)}$ get quantized changes will shrink, and their oscillations around these transition points will also become smaller.}

The topological winding numbers of static SOTIs, as introduced in Sec.~\ref{SSOTI}, could also be extracted from MCDs in a similar manner. A demonstration of this is given in Fig.~\ref{figMCD2}, which corresponds to the SOTI model defined in Eq.~(\ref{mstat}). It is clear that the MCDs averaged over a long time duration $\tau=400$ (triangles in Fig.~\ref{figMCD2}) are consistent with the theoretical values of $\nu_x\nu_y/4$ (solid lines), as predicted by the general relation Eq.~(\ref{CBarAlpha}).

In previous studies, the MCD has been measured in 1D photonic~\cite{MCD2} and
cold atom~\cite{MCD3} systems. An experimental proposal for detecting the MCDs
of $U_{x}^{(\alpha)}$ by implementing quantum walks in BECs has also
been discussed \cite{LW}. According to the tensor product structure
of $U_{{\cal H}}^{(\alpha)}$, one may implement
the dynamics governed by $U_{x}^{(\alpha)}$ and $U_{y}^{(\alpha)}$
in two decoupled 1D systems separately following the proposal of Ref.~\cite{LW},
thereby detecting the topological invariants $\nu_{0,y}$ and $\nu_{\pi,y}$
of Floquet SOTIs introduced in Sec.~\ref{flsoti}.

\section{Concluding Remarks}
\label{conc}

In this paper, we report a theoretical proposal for constructing static and Floquet SOTI by stacking 1D topological phases and coupling them with dimerized hopping amplitude. The total Hamiltonian can then be written as a Kronecker sum of two 1D Hamiltonians describing a static 1D SSH model in the $y$-direction and another 1D static or Floquet topological insulating model in the $x$-direction, allowing one to characterize the existence of the topological corner modes of the whole system by separately analyzing the topology of the 1D model in the $x$- and $y$-directions. 

Although the explicit models presented in this paper possess all inversion, time-reversal, particle-hole, and chiral symmetries, their topological corner modes are protected solely by the chiral symmetry alone, as demonstrated by our numerical results in the presence of perturbations breaking all but the chiral symmetry. The SOTI proposed in this paper is thus fundamentally different from most other existing SOTI proposals, which rely on the presence of spatial symmetries. It is also expected that our proposal can be generalized to a class of 2D systems whose Hamiltonian can be broken down into a Kronecker sum of two 1D Hamiltonian describing any static and/or Floquet topological phases. Moreover, we have also numerically verified that the presence of small perturbations breaking the Kronecker sum structure of the Hamiltonian does not qualitatively affect the existence of the corner modes, provided such perturbations do not close the bulk or edge gap of the system.

It is expected that our proposal above can be extended to higher dimensional systems for constructing static and Floquet higher-order topological phases, which is left for future studies. To this end, one may start with a Kronecker sum of several 1D and/or 2D static or time-periodic Hamiltonian, then tune each of these Hamiltonian in its topologically nontrivial regime. By a similar mechanism elucidated in Sec.~\ref{general}, the resulting system is expected to host topological corner and/or hinge states at its boundaries.  

Finally, we have demonstrated the capability of Floquet SOTI to host arbitrarily many topological corner modes at quasienergy zero and $\frac{\pi}{T}$, which may find its potential applications in quantum information processing (see e.g. Ref.~\cite{RG6,RG5}). It is expected that there are other interesting and unique features of Floquet SOTI with no static analogue that have not been explored in this paper. Further exploration on the physics of Floquet SOTI and higher-order topological phases is thus imagined to be an interesting aspect to pursue in the future. 

{ \emph{Note added.} {  Recently, we become aware of two recent preprints \cite{FSOTI,FSOTI2} which also discuss a proposal for constructing Floquet SOTI.} These papers however use a similar model as that introduced in Ref.~\cite{HTI1,HTI2}, which does not admit a Kronecker sum structure. As a result, while Floquet zero and $\pi$ corner modes can also coexist in such a model, it is not straightforward to find a parameter a regime in which a desired number of Floquet zero and $\pi$ corner modes emerge.}

\section*{Acknowledgements}
 {J.G. acknowledges support from the Singapore NRF Grant No.~NRF-NRFI2017-04 (WBS No.~R-144-000-378-281) and by the Singapore Ministry of Education Academic Research Fund Tier I (WBS No.~R-144-000-353-112). L.Z. acknowledges support from the Young Talents Project at Ocean University of China (Grant No. 861801013196).}

 {R.W.B. and L.Z. contributed equally to this work.}

\appendix
{   
\section{Analytical derivation of topological corner modes} \label{app}

In the following, we present a detailed derivation of the analytical expressions for the topological corner modes presented in the main text, i.e., Eqs.~(\ref{szero}) and (\ref{fcmode}) for the static and Floquet cases respectively. 

In the static case, due to Kronecker sum structure of the Hamiltonian, we may write its eigenstates in the form of $|0_{(X,Y)}\rangle = |0_X\rangle \otimes |0_Y\rangle$, where $|0_X\rangle$ and $|0_Y\rangle$ are respectively zero energy solutions to $H_x$ and $H_y$ which satisfy $\mathcal{H}=H_x\oplus H_y$. In particular, for the explicit model presented in the main text, i.e., Eq.~(\ref{SSH}), we have 

\begin{eqnarray}
H_x &=& \sum_{i=1}^{N_x} \left\lbrace \left[J_x + (-1)^i \delta J_x\right] |i+1\rangle \langle i | +h.c. \right\rbrace \;, \nonumber \\
H_y &=& \sum_{j=1}^{N_y} \left\lbrace \left[J_y + (-1)^i \delta J_y\right] |j+1\rangle \langle j | +h.c. \right\rbrace \;. \label{split}
\end{eqnarray}

\n Both $|0_X\rangle$ and $|0_Y \rangle$ can then be constructed perturbatively and iteratively as follows. We may start by guessing that $|0_{X=1} \rangle=|1\rangle + |0_{X=1}^{(1)}\rangle$, which gives

\begin{equation}
H_x |0_X\rangle = \left( J_x -\delta J_x\right) |2\rangle + H_x |0_{X=1}^{(1)}\rangle \;. \label{pert1}
\end{equation}

\n We continue by choosing $|0_{X=1}^{(1)}\rangle = -\frac{J_x -\delta J_x}{J_x +\delta J_x} |3\rangle +|0_{X=1}^{(2)}\rangle$, so that $H_x |0_{X=1}^{(1)}\rangle$ cancels the first term of Eq.~(\ref{pert1}) and replaces it with a term $\propto \frac{J_x -\delta J_x}{J_x +\delta J_x}$, i.e.,

\begin{equation}
H_x |0_{X=1}\rangle = -\frac{\left( J_x -\delta J_x\right)^2}{\left( J_x +\delta J_x\right)} |4\rangle + H_x |0_{X=1}^{(2)}\rangle \;. \label{pert2}
\end{equation}

\n Following the above step, we further choose $|0_{X=1}^{(2)}\rangle = \left(\frac{J_x -\delta J_x}{J_x +\delta J_x}\right)^2 |5\rangle +|0_{X=1}^{(3)}\rangle$, which replaces the first term of Eq.~(\ref{pert2}) with a term $\propto \left(\frac{J_x -\delta J_x}{J_x +\delta J_x}\right)^2$. This procedure can be repeated indefinitely to obtain

\begin{equation}
|0_{X=1}\rangle = \sum_{i=1}^{N} \left( -\frac{\mathcal{J}_{x}'}{\mathcal{J}_{x}} \right)^{i-1} |2i-1\rangle +\mathcal{O}\left(\left[\frac{\mathcal{J}_{x}'}{\mathcal{J}_{x}}\right]^{N+1}\right)\;, \label{pert3} 
\end{equation}

\n where $\mathcal{J}_{x}'=J_{x}-\delta J_{x}$ and $\mathcal{J}_{x}=J_{x}+\delta J_{x}$. In the topologically nontrivial regime, i.e., $\mathcal{J}_{x}'< \mathcal{J}_{x}$, with a sufficiently long lattice, the correction term to Eq.~(\ref{pert3}) becomes very small and $|0_{X=1}\rangle$ provides a very good analytical approximation to the zero energy solution to $H_x$. It is also evident from Eq.~(\ref{pert3}) that it is localized near one end of the lattice, whose localization length depends on the ratio $\frac{\mathcal{J}_{x}'}{\mathcal{J}_{x}}$. The same procedures can be applied to find $|0_{X=N_x+2}\rangle$, $|0_{Y=1}\rangle$, and $|0_{Y=N_y+2}\rangle$, so that by recalling that $|0_{(X,Y)}\rangle = |0_X\rangle \otimes |0_Y\rangle$, we readily obtain Eq.~(\ref{szero}) in the main text.  

In the Floquet case, the tensor product structure of the Floquet operator, i.e, Eq.~(\ref{fhoti2}), also allows the decomposition of Floquet zero and $\pi$ corner modes into $|0_{(X,Y)}\rangle=|0_X\rangle \otimes |0_Y\rangle$ and $|\pi_{(X,Y)}\rangle=|\pi_X\rangle \otimes |0_Y\rangle$, where $|0_X\rangle$ and $|\pi_X\rangle$ are quasienergy $0$ and $\frac{\pi}{T}$ solutions to $U_{H_{1D}}$, whereas $|0_Y\rangle$ is a quasienergy $0$ solution to $U_{H_y}$. In particular, for the explicit model studied in the main text, i.e., Eq.~(\ref{fhoti1}), $H_y$ is the same static Hamiltonian considered before in Eq.~(\ref{split}). As such, the same $|0_Y\rangle$ found above applies, and Eq.~(\ref{fcmode}) in the main text immediately follows.

\section{Relation between the polarization and chiral winding number} \label{app2}

Consider a chiral symmetric 1D Hamiltonian given by $H(k)=h_1(k)\sigma_x +h_2(k)\sigma_y$. Its energy eigenvalues and eigenstates can be easily found as $E_\pm=\pm \sqrt{h_1^2+h_2^2}$ and

\begin{equation}
|\pm\rangle = \frac{1}{\sqrt{2}}\left(\begin{array}{c}
1 \\
\pm e^{\mathrm{i}\xi}
\end{array}\right) \;,
\end{equation}

\n where $\tan\xi =\frac{h_2}{h_1}$. The polarization $P_\pm$ is defined as

\begin{eqnarray}
P_\pm &=& \oint \frac{dk}{2\pi \mathrm{i}} \langle \pm |\partial_k  | \pm \rangle \nonumber \\
&=& \oint \frac{dk}{4\pi} \frac{d\xi}{dk} \;. \label{chir}
\end{eqnarray}

\n On the other hand, following Eq.~(\ref{wn}) in the main text, we can define the winding number for such a Hamiltonian as

\begin{eqnarray}
\nu_\pm &=& \frac{1}{2\pi \mathrm{i}} \oint h^{-1}\partial_k h \nonumber \\
&=& \oint \frac{dk}{2\pi} \frac{d\xi}{dk}\;, \label{chir2} 
\end{eqnarray} 

\n where $h=h_1+\mathrm{i}h_2=|h|e^{\mathrm{i}\xi}$. By inspecting Eqs.~(\ref{chir}) and (\ref{chir2}), $\nu_\pm=2P_\pm$ easily follows. 

}


\begin{thebibliography}{99}
\bibitem{note2} As we are working under PBC, we can, without loss of generality, shift the center of the lattice to $x=y=0$ for simplicity.
\end{thebibliography}
\end{document}